\documentclass[manuscript,screen]{acmart}

\AtBeginDocument{%
  \providecommand\BibTeX{{%
    \normalfont B\kern-0.5em{\scshape i\kern-0.25em b}\kern-0.8em\TeX}}}

\usepackage{wrapfig}
\usepackage{pdflscape}
\usepackage{booktabs}
\usepackage[normalem]{ulem} 
\usepackage{ulem}

\usepackage{float}
\usepackage{appendix}

\usepackage{xcolor}
\newcommand{\blue}[1]{\textcolor{blue}{#1}}
\colorlet{blue}{black} 
\renewcommand{\sout}[1]{}

\acmDOI{10.1145/3772318.3790280}
\acmISBN{979-8-4007-2278-3/2026/04}

\begin{document}


\title{Co-Designing Collaborative Generative AI Tools for Freelancers}
\author{Kashif Imteyaz}
\email{imteyaz.k@northeastern.edu}
\affiliation{%
  \institution{Civic AI Lab, Northeastern University}
  \city{Boston}
  \state{Massachusetts}
  \country{USA}
}

\author{Michael Muller}
\email{michael.muller.hci@proton.me}
\affiliation{%
  \institution{IBM Research}
  \city{Cambridge}
  \state{Massachusetts}
  \country{USA}
}

\author{Claudia Flores-Saviaga}
\email{floressaviaga.c@northeastern.edu}
\affiliation{%
  \institution{Civic A.I. Lab, Northeastern University}
  \city{Boston}
  \state{Massachusetts}
  \country{USA}
}

\author{Saiph Savage}
\email{s.savage@northeastern.edu}
\affiliation{%
  \institution{Civic A.I. Lab, Northeastern University}
  \city{Boston}
  \state{Massachusetts}
  \country{USA}
}
\affiliation{%
  \institution{UNAM}
  \city{Mexico City}
  \country{Mexico}
}
\renewcommand{\shortauthors}{Imteyaz et al.}

\begin{teaserfigure}
  \includegraphics[width=\textwidth]{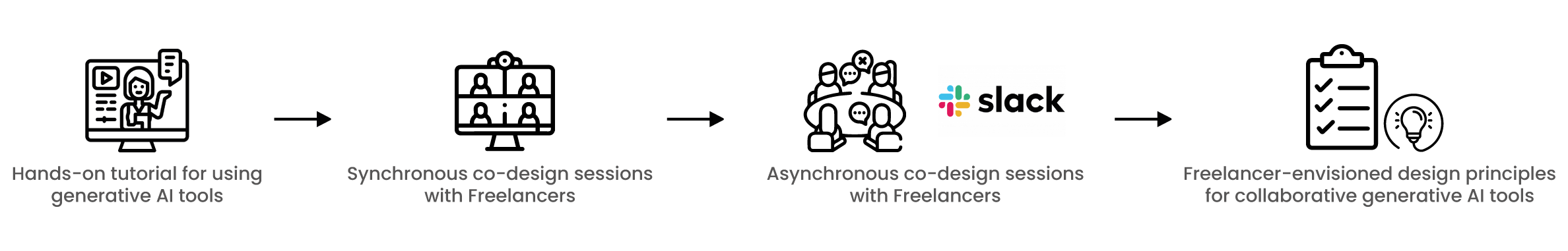}
  \caption{Overview of our study for co-designing collaborative generative AI tools for freelancers.}
  \Description{A flowchart showing four stages of the co-design study: a workshop depicted by a computer screen with four people, followed by co-design insights represented by a lightbulb icon, then a Slack discussion shown by a group of talking people and the Slack logo, and concluding with recommendations symbolized by a clipboard with a checklist. Arrows connect these stages indicating progression from left to right.}
  \label{fig:teaser}
\end{teaserfigure}

\begin{abstract}
Most generative AI tools prioritize individual productivity and personalization, with limited support for collaboration. Designed for traditional workplaces, these tools do not fit freelancers' short-term teams or lack of shared institutional support, which can worsen their isolation and overlook freelancing platform dynamics. This mismatch means that, instead of empowering freelancers, current generative AI tools could reinforce existing precarity and make freelancer collaboration harder. To investigate how to design generative AI tools to support freelancer collaboration, we conducted co-design sessions with 27 freelancers. A key concern that emerged was the risk of AI systems compromising their creative agency and work identities when collaborating, especially when AI tools could reproduce content without attribution, threatening the authenticity and distinctiveness of their collaborative work. Freelancers proposed "auxiliary AI" systems, human-guided tools that support their creative agencies and identities, allowing for flexible freelancer-led collaborations that promote "productive friction". Drawing on Marcuse's concept of technological rationality, we argue that freelancers are resisting one-dimensional, efficiency-driven AI, and instead envisioning technologies that preserve their collective creative agencies. We conclude with design recommendations for collaborative generative AI tools for freelancers.
\end{abstract}



\maketitle
\section{Introduction}
Generative AI tools are rapidly reshaping knowledge work \cite{yun2025generative}. 
A growing number of these systems are designed to enhance individual productivity \cite{dell2023navigating, noy2023experimental, brynjolfsson2025generative}. Most offer limited support for collaboration or collective knowledge-building \cite{kazemitabaar2024codeaid, yun2025generative}. Tools such as ChatGPT enable individuals to brainstorm, draft, and solve problems on their own \cite{song2025beyond}. While collaborative features have recently emerged in tools like ChatGPT Teams \cite{openai_chatgpt_team_2024}, Claude Projects \cite{anthropic_projects_2024}, and GitHub Copilot Workspace \cite{github_copilot_workspace_2024}, most tools and research on collaborative generative AI focus on structured organizational teams \cite{leeimpactCHI2025, johnson2025exploring}. Less is known about how such tools might support freelancers, who work in decentralized, temporary collaborations without organizational infrastructure \cite{jarrahi2020platformic}. Similarly, image generation tools like Midjourney allow the creation of visual content from text prompts \cite{simonen2025initial, thampanichwat2025mindful} but offer limited support for iterative, multi-user refinement \cite{messer2024co}. These examples illustrate a dominant design orientation: generative AI systems are optimized for speed, automation, and individual efficiency rather than for the slower, dialogic, and iterative processes that make collaboration meaningful \cite{woodruff2024knowledge}.


These dynamics can be understood through Herbert Marcuse’s concept of \emph{technological rationality} \cite{marcuse2013one}, which describes the tendency of technology to prioritize efficiency and control over collective engagement and social transformation \cite{yun2025generative, woodruff2024knowledge}. \blue{Most} generative AI tools embody this logic by privileging personal productivity rather than fostering shared understanding or dialogue \cite{liu2024chatgpt, gomez2025human}. As Marcuse warns, such individualization has consequences beyond technical design,\sout{: it reshapes} reshaping professional practices by discouraging mentorship, peer feedback, and collective development of best practices \cite{chiu2006understanding, kellogg2025novice, woodruff2024knowledge}. In the context of generative AI, \sout{technological rationality manifests}\textcolor{blue}{we observe this rationality manifesting} in two ways: (1) discouraging human–human collaboration by reducing spaces for dialogue and peer learning; and (2) narrowing human–AI collaboration into one-dimensional, efficiency-driven assistance rather than dialogic engagement \cite{gomez2025human, komura2026deepening, simkute2025ironies}. 

\blue{ When generative AI tools reinforce this individualistic, productivity-centered logic, they risk narrowing the forms of collaboration that freelancers can imagine or enact. Instead of amplifying the creative, communal, and political potentials within freelance work \cite{kinder2019gig}, the technology can inadvertently reproduce the very one-dimensional structures that limit workers' ability to build shared power and reshape their condition.} \textcolor{blue}{Together, }these reductions risk producing what Marcuse calls \emph{``one-dimensional thinking''} \cite{marcuse2013one}, which discourages critical reflection and limits the imagination of alternative futures \cite{huang2023generative, agarwal2025ai}.

\blue{This efficiency-driven, individualistic orientation of current generative AI tools can be especially harmful for freelancers, who already tend to work in isolation \cite{gray2019ghost, jarrahi2020platformic, jarrahi2021algorithmic}. Unlike traditional employees, freelancers rarely have access to the shared infrastructure, mentorship, and institutional knowledge that organizations use to scaffold collaboration \cite{giacumo2020evidence, stroebaek2013let, waber2010productivity}. As a result, AI systems that focus solely on individual output could further erode the already limited opportunities freelancers have to connect with peers, learn from others, and build collaborative practices.} \blue{Note that maintaining freelancer collaborations is important as these collaborations have produced significant outcomes \cite{valentine2017flash}. Freelancers coordinate diverse teams to create complex creative work \cite{kim2017mechanical, valentine2017flash}, engage in peer learning \cite{margaryan2019comparing, dontcheva2014combining}, form teams for joint bidding \cite{fulker2024cooperation, wang2021examination}, and organize collective actions that challenge platform practices \cite{hsieh2025gig2gether, gray2019ghost}. This tension between structurally enforced isolation and the growing importance of high-stakes collaboration creates a dual challenge: freelancers must sustain human-to-human collaboration under fragile conditions, while most generative AI tools remain designed for individuals rather than teams \cite{woodruff2024knowledge}.}

\blue{Although recent research has begun exploring generative AI for collaboration \cite{he2024ai, mao2024multi, subramonyam2025prototyping}, these efforts predominantly focus on traditional workers in structured organizational settings, overlooking the unique needs of decentralized freelancers who must coordinate temporary teams without institutional support. Without tools explicitly designed for freelancers' context, generative AI risks streamlining individual tasks while failing to support the fragile social arrangements that make freelancer collaboration possible \cite{fulker2024cooperation}.}

\blue{\blue{To move beyond these limitations, we draw on Andrew Feenberg’s concept of \emph{``deep democratization''} \cite{DemocratizingTechnologyAndrewFeenberg}, developed in the tradition of his mentor Herbert Marcuse, which argues that users should play a meaningful role in shaping both the development and the values of the technologies they depend on \cite{friedman2019value}. Guided by this perspective, we conducted a 
series of co-design sessions with 27 freelancers, including developers, designers, writers, project managers, and virtual assistants. Our approach integrates critical theory into the co-design process, enabling participants to question dominant AI logics and explore alternative designs for generative AI supported collaboration. Our study is guided by the following research questions:}}
\begin{itemize}
    \item \textbf{RQ1:} \emph{How do freelancers envision the future of collaborative generative AI tools?}
    \item \textbf{RQ2:} \emph{In what ways can generative AI be designed to enable and strengthen collaborations among freelance knowledge workers?}

\end{itemize}

\sout{Our co-design sessions with 27 freelancers revealed that, through their design visions, they actively resisted technological rationality: the notion that modern technology should prioritize efficiency and control as ultimate values. Instead, they imagined collaborative AI tools that would strengthen human-to-human interaction rather than replace it with human–AI efficiency.}

\textcolor{blue}{Through our co-design sessions, freelancers articulated a vision that actively resists technological rationality by prioritizing human-to-human interaction over human-AI efficiency.} To articulate these futures, freelancers first identified how technological rationality appears in current generative AI tools. They pointed to the lack of situated knowledge, which often leads to generic outputs rather than contextually relevant solutions useful for their \blue{different} collaborations, \blue{including creative co-design, peer learning, and collective action}. They also highlighted the risks of over-reliance on such tools, noting that it can weaken critical thinking and creativity in group work. In addition, freelancers raised ethical concerns, particularly that generative AI tools might compromise creative agency and work identity by reproducing content without attribution, thereby threatening the authenticity and distinctiveness of their collaborative outputs.

Building on these critiques, freelancers envisioned AI as an \emph{auxiliary collaborator}: supportive but not directive, enhancing collaboration while preserving human creativity \blue{and allowing for productive friction.} \sout{direction, and authenticity.} Drawing on Marcuse’s concept of technological rationality \cite{marcuse2013one}, we argue that freelancers resist one-dimensional, efficiency-driven AI and instead envision technologies that preserve collective and creative agency.

\blue{Our work makes three main contributions to HCI research on collaborative generative AI tools for freelancers. We develop these contributions through a co-design approach that puts critical theory into practice using Future Workshops and AI-generated design probes, which help freelancers question and resist dominant AI logics:}

\begin{itemize}
    \item \textbf{Empirical Insights into Current Limitations of Generative AI Tools for Freelancers' Collaboration:} We analyze how \textcolor{blue}{the inherent \textit{technological rationality} of current }\sout{existing} generative AI tools \sout{fall short in supporting} \textcolor{blue}{fails to support} freelancer collaboration. 
    \item \textbf{\blue{Ethical Challenges of Generative AI Tools for Freelancers' Collaboration}}: \blue{We identify the ethical challenges freelancers face when incorporating generative AI into their collaborative work, particularly how these tools can undermine their collective creative agency and weaken their work identities}.
   
    \item \textbf{Design Recommendations for Generative AI Tools for \blue{Freelancer} Collaboration}: 
    \blue{We propose designing collaborative generative AI tools 
    that treat AI as an auxiliary collaborator and support “productive friction''.}
\end{itemize}

\section{Related Work}
\subsection{\blue{Freelancers and Collaborative Work}}
\blue{We define freelancers as self-employed knowledge workers who contract with multiple clients on digital labor platforms (e.g., Upwork, Freelancer.com, Fiverr) rather than through traditional employment \cite{gandini2016digital}. Their access to work is shaped by platform algorithms \cite{wood2019good}, and they typically perform knowledge-intensive tasks such as software development, design, writing, and consulting \cite{fulker2024cooperation}.} \blue{Unlike remote employees, who retain continuous employment and access to company tools, training, and HR support \cite{malone1994interdisciplinary,devadoss2007enterprise}, freelancers must secure their own clients, provide their own infrastructure, and absorb risks that organizations usually bear \cite{wood2019good,wilkins2022gigified}. Lacking shared software, workflows, and institutional knowledge \cite{retelny2014expert,kim2017mechanical,luther2014crowdcrit}, freelancers nonetheless represent a substantial segment of the workforce: over 1.5 billion people worldwide engage in freelance work, projected to comprise roughly one-third of the global workforce \cite{Upwork2025FutureWorkforceIndex, Statista2025GigFreelancersUS, Statista2025_GlobalFreelancerTalentPools}. Studying freelancers is therefore essential for understanding the future of work.}

\subsubsection{\blue{Defining the Nature of Collaboration in Freelance Work}}
\blue{In this section, we outline how freelancers collaborate in practice, since these dynamics are essential for interpreting the generative AI tools they envisioned and their reflections on existing systems.} We define collaboration in freelancing as situations where two or more freelancers coordinate to produce outcomes that would be impossible or substantially harder to achieve alone \cite{retelny2014expert,kim2017mechanical,luther2014crowdcrit}. For example, a writer and graphic designer partnering on a marketing campaign, aligning timelines, dividing roles, and integrating outputs across tools and platforms.  Because freelancing lacks stable organizational structures, workers must assemble temporary arrangements for each project \cite{jarrahi2021flexible}, deciding how to communicate, divide tasks, and resolve conflicts in contexts where collaborators may also be competitors \cite{wood2019good,wilkins2022gigified}. Although some have questioned whether meaningful collaboration is possible under such conditions, recent work reports cooperation rates of about 85\% among Upwork freelancers \cite{fulker2024cooperation}, suggesting that freelancers can achieve teamwork even without the organizational supports available to traditional workers \cite{retelny2014expert,kim2017mechanical,luther2014crowdcrit}.

Freelancer collaboration takes multiple forms, including peer learning through informal networks \cite{gray2019ghost}, subcontracting to fill skill gaps \cite{morris2017subcontracting, kittur2013future}, joint bidding on projects requiring diverse expertise \cite{wang2020crowdsourcing, wang2021examination}, creative co-production in design and media \cite{oppenlaender2020creativity, valentine2017flash}, and collective action through tools \cite{imteyaz2024gigsense, hsieh2025gig2gether}. We provide detailed descriptions of each collaboration type in Appendix~\ref{appendix:collab-types}. However, this work has not yet grappled with generative AI. The intersection of freelancers’ collaboration needs and generative AI capabilities remains largely unexplored, particularly for supporting temporary, trust-based collaboration while preserving creative authenticity.

\subsection{Collaborative Generative AI Tools: Existing Approaches and Gaps}
\textcolor{blue}{Prevailing human–AI collaboration research emphasizes shared understanding and communication rather than full automation \cite{amershi2019guidelines, bansal2024challenges}. These frameworks typically focus on individual decision-making, where AI offers recommendations and humans retain oversight \cite{gomez2025human, amershi2019guidelines}. While much of this work centers on single users working with AI assistants \cite{zhang2021ideal, momose2025human}, recent studies have begun to examine collaborative uses of generative AI tools.}

\textcolor{blue}{Researchers have examined how people use generative AI while collaborating in design and ideation \cite{johnson2025exploring}, qualitative analysis \cite{gao2023coaicoder}, and cooperative games \cite{sidji2024human}. Other work investigates multi-user interaction with ChatGPT or other LLM-based tools for co-located brainwriting \cite{shaer2024ai}, group ideation \cite{he2024ai}, collaborative music composition \cite{suh2021ai}, LLM-driven shared displays for team ideation \cite{zhang2025ladica}, and LLM-based tools that act as counter-arguers in group decision-making \cite{LLMasDeviladvocate}. Recent research on generative AI in group contexts further shows that AI can both support and undermine collaboration, sometimes improving information sharing and coordination, and other times fostering over-reliance that weakens critical discussion \cite{zhang2021ideal, houde2025controlling, leeimpactCHI2025, simkute2025ironies}. However, this literature largely focuses on teams in structured organizational settings with shared infrastructure and relatively stable roles. As a result, there is still a research gap around generative AI for freelancer collaborations, where teams are temporary, distributed across platforms, and governed by delicate contracts \cite{gray2019ghost}. It is not evident that findings from organizational teams transfer to freelancers, whose collaborations must accommodate fluid membership, uneven technical skills, and fragile client relationships \cite{sutherland2017gig}.}

\textcolor{blue}{Commercial Generative AI providers now offer collaborative features, such as shared conversations in ChatGPT Teams \cite{openai_chatgpt_team_2024}, persistent context in Claude Projects \cite{anthropic_projects_2024}, and multi-developer support in GitHub Copilot Workspace \cite{github_copilot_workspace_2024}.} \blue{Beyond these commercial platforms, emerging generative AI tools enable teams to work collectively with AI for ideation, co-creation, and decision-making \cite{seymour2024speculating, shneiderman2022human, johnson2025exploring}. However, these tools remain largely designed for traditional organizational teams with stable structures, defined roles, and consistent technical infrastructure \cite{johnson2025exploring}. Studies also note that many generative AI systems require explicit prompting, which introduces additional overhead \cite{johnson2025exploring, cogniChallangeinPrompt}, a challenge that may be especially acute for freelancers managing fragmented, time-constrained collaborations \cite{graham2019global}. It remains unclear whether these approaches meet the needs of decentralized freelancers coordinating temporary teams across platforms without organizational support \cite{jarrahi2020platformic}.}



\subsection{Co-Design Approaches}
Traditional design often follows a linear path from problem identification to solution generation \cite{norman2013design}, which can lead to premature solutions that overlook nuanced user needs \cite{thyme2021discovery,costanza2021design,norman2013design,buchanan1992wicked} and fail to address deeper issues \cite{sanders2014probes}. Participatory and co-design approaches emerged to counter these limitations \cite{niedderer2020working}. They draw on diverse theories and practices \cite{halskov2015diversity,roberts2022datafication,muller2012participatory} and aim to democratize design, address power relations, foster mutual learning, and explore alternative visions \cite{elsayed2023exploring,luck2018makes}. Despite differences and practical constraints across methods \cite{ehn2014making, greenbaum2020design, muller2012participatory, simonsen2013routledge}, we treat this diversity as a resource for tailoring approaches to specific design contexts.

\subsubsection{Remote Participatory and Co-Design Methods}
Freelancing platforms connect geographically dispersed workers and employers, making remote participatory and co-design sessions essential \cite{he2023understanding, leporini2022making, maus2021designing}. Although online co-design methods predate COVID-19, their use expanded rapidly during the pandemic \cite{farrell2006pictiol, walsh2015case}. Prior work documents challenges such as participation barriers, logistics, and privacy concerns \cite{dalsgaard2022four, elsayed2023exploring, lehdonvirta2018flexibility}, but also successful implementations, especially with remote-savvy participants \cite{he2023understanding, muller2012brainstorming, sun2022investigating}. Our study engages freelancers in co-designing generative AI tools for collaboration, drawing on design probes \cite{zhang2023stakeholder}, speculative job design \cite{ma2025speculative}, and structured remote methods \cite{sun2022investigating, steen2013co}. To mitigate limits of remote interaction \cite{robertson2012challenges}, we used flexible engagement \cite{bjorgvinsson2008open}, consistent interaction \cite{o2004participatory}, relationship-building \cite{slingerland2022participatory}, shared knowledge repositories \cite{pilemalm2008third, imteyaz2024gigsense}, structured facilitation \cite{constantin2021distributing}, and iterative feedback over several months to sustain co-design against technologically driven isolation.

\subsubsection{Design Probes using DALL-E Images}
Within participatory and co-design approaches, design probes are used to elicit creative reflection and expression. As “directed craft objects,” they help participants surface personal experiences and translate them into design ideas \cite[p.~1]{walsh2016inclusive}. \blue{Recent work has begun to explore generative AI as a creativity support tool for individual designers \cite{cai2023designaid, wang2025exploring} and for speculating about climate futures \cite{lc2023speculative}.} We extend this tradition by using DALL·E to co-create visual probes with freelancers. These AI-generated images help participants externalize and reflect on their visions for \textcolor{blue}{their} collaborative AI tools. Building on \citet{designprobeswallace2013making} and \citet{williams2002self}, we use probes to reduce self-censorship and scaffold creative expression. Through collective interpretation of the images, freelancers discuss complex, often abstract questions about AI collaboration. Like traditional probes, these visuals act as “sites and tools for reflection” \cite{designprobeswallace2013making, cai2023designaid}, mediating dialogue and surfacing aspirations and concerns. Responding to critiques that reflective design can be difficult to operationalize \cite{ChallangesinCriticalDesign}, our method embraces ambiguity and uses AI-generated probes to help freelancers imagine alternatives to dominant technological paradigms.

\subsection{Critical Theory and Technology Design}
\subsubsection{Marcuse's Critique of Technological Rationality}
Herbert Marcuse’s concept of technological rationality describes how modern technological systems embed an ideology of efficiency and control that flattens complex human realities into standardized outputs \cite{marcuse2013one, bardzell2018critical}. In One-Dimensional Man, he argues that this produces “one-dimensional thinking,” where critical reflection and the ability to imagine alternative social or technological orders are absorbed into dominant economic imperatives \cite{to2023flourishing}. In the context of generative AI, this critique helps explain how such systems can impose standardized approaches to creative work that undermine freelancers’ contextual and collaborative labor. As \citet{christian2014digital} notes, digital labor platforms already enact technological rationality by reducing complex skills to quantifiable metrics, with consequences that include diminished critical thinking and growing technological dependency.

\subsubsection{Critical Perspectives on AI and Platform Labor}
Recent critical work on AI and platform labor shows how these systems extract value while obscuring worker contributions \cite{crawford2021atlas, fan2020crowdco}, and how scale often conflicts with meaningful participation \cite{kang2022stories, young2024participation, do2024designing, imteyaz2024gigsense}. While Marcuse offers a foundational critique, other perspectives extend this lens. Feminist theories of situated knowledge \cite{haraway2013situated} and critical computing \cite{bardzell2018critical} show how AI can reinforce dominant norms, empirical studies examine how designers navigate ethical tensions \cite{Chivukula}, and algorithmic management research documents how complex labor is reduced to standardized metrics \cite{gray2019ghost, jarrahi2021algorithmic}. Marcuse also emphasized that technology can support human liberation if redesigned around alternative values \cite{marcuse2013one}. This tension is salient for generative AI in freelance work, which offers new capabilities while potentially constraining creative autonomy \cite{huang2024design, imteyaz2024human}. Building on this tradition, our study investigates how freelancers resist technological rationality through co-design, surfacing alternative values to navigate stakeholder tensions \cite{gray2019ghost, scholz2017uberworked, friedman2019value}. Unlike prior work that primarily critiques AI or documents freelancer experiences \cite{kinder2019gig, do2024designing, ChallangesinCriticalDesign, OneAIdoesNotFitAll}, we integrate critical theory directly into the co-design process \cite{o2004participatory, marcuse2013one} to actively challenge dominant AI logics.

\section{Methods}
Our IRB-approved methodology (Figure \ref{fig:teaser}) combined synchronous and asynchronous co-design. We conducted Zoom sessions with 27 freelancers (groups of 4–5) between April and May 2024, followed by a three-month Slack-based co-design phase with all 27 participants from June to August 2024. Participants received \$20 for the synchronous session, \$10 for the asynchronous phase, plus engagement bonuses. We recorded and transcribed all discussions, collected visual design probes, and conducted qualitative analysis on the resulting data.

\subsection{Procedure for Synchronous Co-Design Sessions}
Each synchronous co-design session included 4–5 freelancers and a moderator from our research team. To prepare participants to design generative AI collaboration tools, we began with a short tutorial on generative AI, then used the “Future Workshops” participatory design method \cite{kensing2020generating}, which moves through critique of the present, envisioning the future, and planning the transition. In total, we conducted 7 synchronous sessions.

\subsubsection{Generative AI Tutorial}
This tutorial familiarized participants with generative AI by: (1) introducing common tools; (2) demonstrating ChatGPT for text generation and DALL·E for image creation; and (3) providing hands-on practice so participants could explore the tools’ capabilities and limitations. \textcolor{blue}{We used DALL·E as a co-design probe to help participants externalize and visualize abstract collaborative futures they struggled to express verbally \cite{williams2002self, zamenopoulos2021types}. Its iterative, text-based interface allowed freelancers with diverse technical backgrounds (see Appendix Table \ref{tab:participants} for participants’ backgrounds) to quickly generate, refine, and discuss visual representations of AI-mediated collaboration, lowering barriers to visual expression and grounding discussion in shared artifacts \cite{zamenopoulos2021types}.} We then proceeded with the three phases of the Future Workshops methodology \cite{kensing2020generating}, with each phase’s outputs directly informing the next.

\subsubsection{Co-Design Phase I: Critiquing the Present Challenges of Generative AI Tools for Collaboration} In this phase, freelancers identified key challenges in using generative AI tools for collaboration, reflecting on frustrations, unmet needs, and design gaps. \textcolor{blue}{The goal was to document current collaboration problems with generative AI, producing pain point analyses and critical visual probes to ground Phase II’s visioning activities.} To support expression, we provided access to DALL·E. Participants generated images representing their experiences, which served as design probes to spark reflection and discussion \cite{designprobeswallace2013making}. This approach can reduce creative barriers and enable more open expression \cite{hunter2024creativity, williams2002self}. Sharing these visuals with peers helped surface recurring pain points in current AI tools for collaboration. \textcolor{blue}{Appendix~\ref{appendix:appen} presents example design probes from our co-design sessions.}



\subsubsection{Co-Design Phase II: Envisioning the Future of Generative AI Tools for Collaboration}
In the second phase, we invited participants to envision future generative AI tools for collaboration. \textcolor{blue}{Building on Phase I’s challenges, this phase focused on imagining ideal tools that could address those pain points, producing speculative design concepts and aspirational visual probes.} The moderator prompted speculative thinking about desired functionalities, collaborative scenarios, how freelancers would use these tools in practice, and potential real-world challenges. Participants again used DALL·E to visualize their ideal tools, creating design probes that turned abstract ideas into concrete concepts and sparked discussion \cite{designprobeswallace2013making}. They then presented their visuals, received peer feedback, and iteratively refined their designs. This process encouraged them to move beyond incremental fixes and imagine more transformative futures for AI-supported freelance collaboration, surfacing key values, features, and concerns.

\subsubsection{Co-Design Phase III: Transitioning from Challenges to the Future of AI Collaboration}
In the final phase, participants focused on bridging the gap between current challenges and their envisioned futures for AI-supported collaboration. \textcolor{blue}{Synthesizing insights from Phases I and II, this phase aimed to map concrete pathways from current problems to desired futures, producing implementation strategies and transition-focused visual probes.} Freelancers identified actionable steps, innovations, and interventions needed to realize their visions. Using DALL·E, they generated images illustrating potential future directions for their proposed tools. These served as design probes \cite{designprobeswallace2013making}, supporting clearer articulation of ideas and enabling concrete discussion of how to make these tools real. Moderators guided the transition-focused discussion with prompts about immediate next steps and potential implementation barriers. Participants shared pathways and visuals, received peer feedback, and refined their ideas through structured discussion. At the end of each phase, we compiled the visual design probes and discussion transcripts.

\subsection{Procedure for Asynchronous Co-Design Session}
Following the synchronous co-design sessions, we had a three-month asynchronous phase on Slack with all 27 freelancers. This let participants revisit and expand their ideas, engage with proposals from other groups, and contribute on their own schedules and time zones, supporting broader inclusion. It also enabled deeper reflection and iteration, allowing freelancers to explore design challenges more thoroughly and synthesize ideas across sessions.

\subsubsection{Structure}  
To promote broad participation and collaborative refinement, we structured the asynchronous phase around a shared Slack workspace where freelancers could engage at their own pace. A moderator summarized key insights from the synchronous sessions, including: (1) core challenges with current generative AI tools for collaboration, (2) co-created visions for AI-supported collaboration, and (3) strategies for realizing those visions. The moderator also shared a Google Drive folder with all AI-generated images, which served as design probes. We tracked which freelancers contributed to specific ideas and invited them to clarify or expand on their proposals, tagging additional freelancers as concepts evolved. Participants were asked to provide feedback on strengths and weaknesses, offer alternative perspectives, and suggest improvements. This iterative, peer-driven process enabled all freelancers to shape each design idea and fostered a sense of collective ownership over the future of generative AI tools for collaboration.

\subsubsection{Final Outcomes of the Co-Design Sessions} 
By the end of the asynchronous phase, we compiled: (a) challenges with current collaborative AI tools \textcolor{blue}{for freelancers}, illustrated with AI-generated visuals; (b) freelancer-envisioned generative AI tools supported by visual probes and descriptions; and (c) strategies for transitioning toward these futures. In our Appendix \ref{appeni}, we present how different design probes evolved through the co-design sessions.

\subsection{Data Analysis}
Drawing on reflexive thematic analysis \cite{terry2017thematic, braun2019reflecting}, we conducted a bottom-up analysis of data from our synchronous and asynchronous co-design sessions. Two researchers independently open-coded separate halves of the transcripts with attention to our research questions on collaborative generative AI design for freelancers, meeting weekly to discuss emerging patterns, refine the codebook, and resolve disagreements through consensus. \textcolor{blue}{Consistent with \citet{braun2019reflecting}, we did not compute inter-rater reliability, focusing instead on interpretive depth and reflexive engagement.} After open coding, we iteratively grouped codes into higher-level themes through axial coding, examining relationships between codes and their relevance to freelancers’ visions of collaborative generative AI tools. \textcolor{blue}{We continued analysis until reaching what Braun and Clarke describe as “meaning sufficiency” \cite{braun2019reflecting}, where no substantially new insights emerged and themes were well supported across the data.} This process produced 18 axial codes, which we synthesized into three core themes that structure our findings. Throughout, we maintained reflexive memos documenting analytical decisions. Participant quotes are anonymized as “P” followed by a participant ID.

\subsection{Participants}
We recruited participants through a job listing on Upwork, attracting 111 freelancers. Upwork was chosen for its broad use \cite{kassi2018online, gandini2016reputation} and ability to surface diverse perspectives \cite{wood2018workers}. A pre-survey screened for eligibility: participants had to be at least 18 years old, have a minimum of one year of freelancing experience, and demonstrate proficiency in English. We then selected 27 freelancers (11 women, 16 men), aged 18–39 (Mean = 29.92, SD = 5.737). Table \ref{tab:participants} in our Appendix summarizes their demographics. \blue{Participants included technical workers (8 developers/data professionals), creatives (7 designers/writers/editors), business service providers (8 in HR/project management/marketing), and support roles (4 virtual assistants/translators/legal). This diversity reflected varied generative AI use cases, from code generation to visual design and writing, and many  collaborated with other freelancers on distributed teams and joint projects.}

\section{Results}
We present the core themes that emerged from our qualitative analysis, organized into four subsections: 

\subsection{The Logic of Efficiency: Technological Rationality in Generative AI for Freelancer Collaboration}
Our co-design sessions demonstrated how current generative AI tools embody Marcuse’s notion of technological rationality, which prioritizes standardized efficiency over human agency. This was clear as freelancers explained how current generative AI tools usually undermine their collaborations by imposing uniform responses and reducing complex human realities to one-dimensional technical problems.

\subsubsection{Generative AI Fails to Understand Freelancers’ Collaborative Context} 
\sout{Freelancers reported that current generative AI tools often lack awareness of their unique work contexts, producing outputs that are too generic, misaligned with their needs, or outright incorrect. Unlike structured and traditional workplaces, freelance collaborations can be dynamic, non-standardized, and usually dependent on interpersonal coordination \cite{fulker2024cooperation, wood2019good}.} 

\sout{This tension between AI's standardized responses and freelancers' situated knowledge reflects what Marcuse identified as technological rationality's reduction of complex human realities to uniform, manageable inputs \cite{marcuse2013one}.}

\blue{Freelancers reported that current generative AI tools rarely understand their work contexts, often producing generic or misaligned outputs. Participants attributed this to AI tools being designed for traditional workplace structures rather than the flexible, temporary collaborations that characterize freelance work \cite{wood2019good}}. \blue{For example, P16 described collaborating with freelancers in fields such as medicine and law on a creative co-production project. In these kinds of collaborations, it is common for freelancers from different disciplines to work together \cite{fulker2024cooperation}, since freelance collaborations are often less formally structured and have fewer institutional barriers to cross-sector participation, which can lead to highly diverse teams \cite{valentine2017flash}. However, the AI did not account for this cross-disciplinary teamwork and failed to understand this collaborative context. Instead, it provided standardized, generic responses that ignored the specific needs and context of their highly specialized collaboration: \emph{``I had to help with a medical school presentation, they were going to pitch to their physician and their attorney [...] I thought I'd take help of AI. I'd give the presentation to the AI and ask the AI questions [...] The limitations I found out [the AI’s limitations in responding to the freelancer’s collaboration-related questions about the presentation] were the ones like, AI doesn't get the contextual meaning. What's the conversation you are having? Like, it doesn't understand the context. And second, which I understand is, okay, there is no intelligence. Like, if you'd ask a question from a fellow human and they'd give you an unexpected answer, you'd switch your question according to that answer. Okay, but AI would not. AI would, you know, keep going on with the script.''} Such experience reveals that when multiple freelancers are coordinating across different disciplines, the AI still responds as if it were addressing a single user. The AI could not understand the shared task, the shifting roles involved, which can also have people working asynchronously, or the iterative adjustments collaborators make as they negotiate meaning together.}

\blue{P22 similarly reported that generative AI outputs were not relevant to the context of their collaborations, which limited the usefulness of these tools. In particular, they wanted to use generative AI to help them teach other freelancers and support peer learning. However, they felt that the examples the AI produced did not reflect the regional realities of freelancers, making those examples feel unrelatable and less useful: \emph{"maybe it is especially because I like to teach others [participating in peer learning], so when I want relatable examples [relatable teaching examples], whatever it's [AI is] giving me, it's [AI-generated content is] not within their locality [the AI fails to generate content relevant to the freelancers’ regional backgrounds]. Normally, I need to still go back and maybe research, do my own [teaching examples for other freelancers], figuring out the right examples that I can use with my learners [freelancers participating in peer learning] that are relatable [...] It [AI] can be giving me examples, sure. But, they [the examples the AI gives] are not relatable to things within their locality or within their area. That is, things that can be well understood by the learners, 
who can relate to it.''} This mismatch highlights a deeper issue: peer learning depends on examples that emerge from a group's shared background and situated knowledge, yet the AI flattened the group's local, relational knowledge into generic outputs that erased the complexity of their collaborative learning environment.} \blue{Along similar lines, P11 hoped to use generative AI to support their collective action. For instance, to help identify how they could collaboratively address the problems they faced. However, they found that the AI’s responses were too generic to be useful in real freelancer collaborations: \textit{"...the AI [output of the generative AI tool] is very generic. We [when collaborating with each other] need more specific, you know, solutions 
to our problems [collective problems of freelancers]. That's one of the issues [with generative AI tools]."} This frustration highlights that generative AI likely cannot represent problems originating from a collective or offer solutions relevant to shared experiences. The AI defaulted to decontextualized responses that ignored freelancers' collective dimension.} \blue{Taken together, these accounts highlight a tension between AI's standardized responses and freelancers' situated knowledge. This tension echoes Marcuse’s critique of technological rationality, where complex human realities are reduced to uniform, manageable inputs \cite{marcuse2013one}.}

To further explain the limitations of generative AI, freelancers compared AI-generated support to responses from online freelancer communities, such as Reddit, where people with firsthand experience can offer more contextually relevant solutions. For example, P9 also liked to participate in peer learning and relied on other freelancers for support when learning how to solve specific problems on the job. P9 felt that the explanations they received from peers were more helpful than the generic responses provided by generative AI: \textit{"Reddit is one of my favorite community apps to use. And, like, the thing with Reddit is most of the answers on Reddit, they are given by real people, they have faced these problems in real time. So the success rate in those situations is far higher than ChatGPT [Generative AI tool]. They [freelancers collaborating with each other] have faced this problem and they are able to solve it. So there's a very high chance that I might be able to solve it, if I'm facing their same problem. So that's the difference between these kind of forums and AI."} This comparison reveals a distinction: freelancers rely on the lived experiences of their peers, who understand the same work context, while AI can provide decontextualized responses that lack this situated knowledge.

\sout{Similarly, P11 noted that AI-generated responses tend to be too generic to be useful in real freelancer collaborations: "...the AI [output of the generative AI tool] is very generic. We [freelancers collaborating with each other] need more specific, you know, solutions [outputs] to our problems [collective problems of freelancers]. That's one of the issues [with generative AI tools]." Such a critique that "the AI is very generic" while freelancers need "specific solutions" illustrates how technological rationality's universalizing logic conflicts with the contextual understanding essential for collaborative work.}

Freelancers also felt that AI’s lack of contextual understanding often produced misleading or counterproductive results, wasting their time instead of improving efficiency. P2 described this frustration when he had tried using generative AI to support collective action, especially to understand freelancers' collective challenges and identify solutions to the problems: \textit{"...I put out my message [prompt to a generative AI tool] that this is what I'm expecting. But most of the times what I've seen in ChatGPT is like it assumes something on its own. It generates something which is not very useful. And when that happens, then it's a waste of time. The whole purpose of ChatGPT is to help us in giving quicker solutions [solutions to their collective problems], but if it [generative AI] interprets it [their collective problem] in a different way [different context], or sometimes it thinks that the problem statement [problems freelancers are facing collectively] is more complex. But actually, the problem statement remains very simple. So in those kind of scenarios, it [AI] gives totally opposite results of what we 
are expecting and it leads to lot of time waste."}  

\begin{wrapfigure}{r}{0.5\columnwidth}
    \centering
    \vspace{-10pt} 
    \includegraphics[width=0.48\columnwidth]{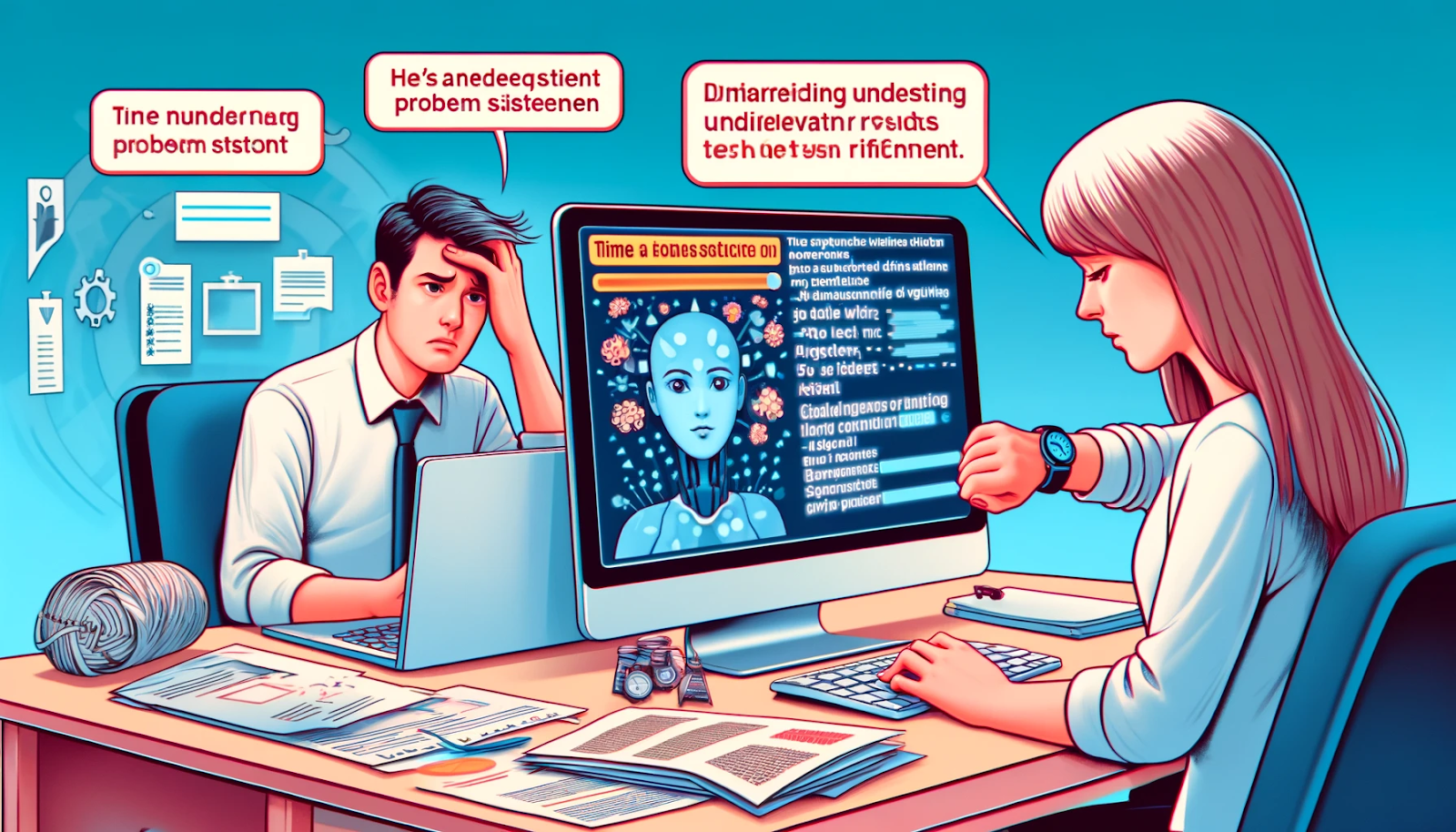}
    \caption{Design probe from Phase I showing how generative AI wastes freelancers' time with irrelevant, homogenized outputs.}
    \label{fig:NoContext}
    \vspace{-10pt} 
\end{wrapfigure}

Fig. \ref{fig:NoContext} shows a visual probe that our participants created to illustrate how generative AI’s lack of context awareness can lead to wasted time \textcolor{blue}{during freelancers’ collaborations}. These collaborative experiences highlight the false promise of technological rationality: although AI systems are designed to increase efficiency, in practice they can create counterproductive work and undermine the very goals they are meant to support.

\subsubsection{Over-Reliance on AI Can Undermine Freelancers’ Creativity and Decision-Making in Collaborative Projects}

\blue{As noted earlier, freelancers frequently engage in collaborative creative work \cite{retelny2014expert,gandini2016reputation,wood2019good}. This might involve completing tasks involving creativity and turning to peer networks for support \cite{margaryan2016understanding}. For instance, P18 described their process of seeking help with video editing: \emph{``when I run into issues [in my freelancing job as a video editor], I mostly ask my friends [freelancer friends] first, because some of my friends have a lot of experience on video editing. But after that, if my friends don't give me some answers that I want, I often find some answers on Discord, which is a platform that video editors [freelancers] use a lot. I just go to some server, just some Q\&A section, and ask a question...''}}

\blue{Building on these existing practices, participants envisioned how generative AI could further support joint creative work. P17, for example, imagined AI acting as a partner in co-designing app interfaces:
\emph{``...if there is a specific, you know, a specific problem [design problem] that we [group of freelancers collaborating] are trying to solve, and let's say that we are trying to solve it through the app interface [solve the problem through the co-design of the app interface]. And to solve this UI issue, I think AI can also be helpful [...] in the future, AI would focus more on creating interfaces that are very intuitive or very creative and, you know, something that really changes or, like, something that really focuses on the entire experience of the user.''}}

\begin{wrapfigure}{r}{0.5\columnwidth}
    \centering
    \vspace{-8pt}
    \includegraphics[width=0.48\columnwidth]{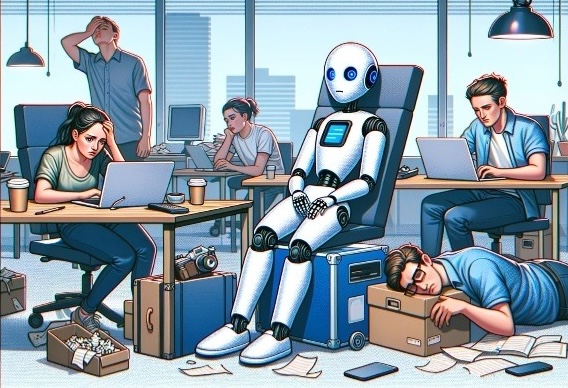}
    \caption{Phase I probe showing AI-induced over-reliance.}
    \label{fig:SkillDetor}
    \vspace{-10pt}
\end{wrapfigure}

\blue{However, this excitement about AI supporting their creative collaborative work was tempered by concerns about how generative AI might reshape freelancers' collaborative dynamics. In particular, participants worried that relying too heavily on generative AI tools could erode their teams' critical thinking, creativity, and independent decision-making. P10 observed how this reliance was already affecting the quality of the creative content that freelancers generated collaboratively: \textit{"They [freelancers] all are producing the same content [when using generative AI tools]...They aren't even trying to learn. So an undesirable use [of generative AI tools] would be that humans would just stop trying to learn, trying to be creative."}}

\blue{Fig. \ref{fig:SkillDetor} shows a design probe illustrating how, during collaboration, freelancers who become over-reliant on generative AI may grow lazy and disengaged instead of actively contributing. Even when the AI is not helping the team’s productivity, this reliance can lead freelancers to slack off rather than move the collaboration forward.}

\blue{Technological rationality is often criticized for its homogenizing effect on creativity \cite{marcuse2013one}. When teams using a system such as generative AI all produce similar outputs and show little inclination to innovate, this pattern suggests that the technology may be flattening diverse forms of expression into more uniform results \cite{marcuse2014aesthetic}.} \textcolor{blue}{Furthermore, participants observed that this "habitual dependency" can extend to minor tasks within a collaboration, potentially causing skills to degrade. } \textcolor{blue}{Similarly, P20 described how this dynamic can discourage freelancers from relying on their own capabilities when working together: \textit{"It's like ChatGPT [Generative AI tool] makes us dependent on ChatGPT. What are we [freelancers collaborating] supposed to do? You know, there are many tasks that actually we 
were supposed to do. But, it 
is creating dependencies, so much that we rely on ChatGPT obviously right now, as well as for some other things..."} }

\blue{Such habitual dependency suggests a subtle erosion of collaborative capacity: when freelancers instinctively turn to AI for even minor tasks, they may gradually lose confidence in each other's skills and judgment during joint work.} \blue{P22, who specializes in DevOps and cloud infrastructure, shared how generative AI powered tools had begun replacing the critical reasoning process of him and his freelance team, particularly when working with his team to create deployment pipelines: \textit{"[Should] we also say that we [he and his freelance team] are over-relying on this [generative AI tools]? Absolutely. An example is me. Every other task that I get before even I think of maybe ways of solving this particular task, I must first see: Maybe I can test that question [task] in GPT [Generative AI tool]. It 
gives me its own reasoning. That means I'm no longer relying on my own sense before making a particular decision. I feel like it is creating an over-reliance on these AI tools."}} 

\blue{What emerges here is a concerning shift in collaborative practice: rather than teams reasoning through problems together, AI seems to become the first source of analysis, potentially displacing the collective deliberation that builds shared understanding.} \blue{Beyond concerns about skill erosion, P4 also emphasized that relying too much on generative AI could lead freelancers to accept AI outputs without questioning them. In collaborative settings, this could mean treating the AI’s response as a final answer instead of using it as a starting point for peer discussion and collective decision-making: \textit{"People [freelancers collaborating] over-rely on AI [generative AI tools] these days. Like you ask someone [a freelancer you are working with], how do you know whether an information the AI gave us [team of freelancers] is correct? So you see, some people don't even check it [AI generated outputs], they just go, they are not creative..."}}

\blue{In light of these risks, freelancers emphasized that generative AI should support, not replace, human judgment in creative work. They argued that it is especially important to preserve space for human expertise and contextual understanding when making high-stakes collective decisions. For example, P5 stressed that when deciding which freelancers to recruit for a joint bid on a creative co-creation project, a human should always make the final choice. This is a high-stakes decision because the team’s composition can affect their chances of winning the project and, ultimately, the livelihoods of everyone involved. For this reason, participants believed that humans must remain directly involved in making such decisions: \textit{"One undesirable, uh, thing with AI [generative AI tools] is, you know, this sort of over-reliance on the, on the system [...] currently, with the level at which we, the level at which it [generative AI tools] works, it's not advisable for us to place 100\% reliance on it [generative AI tools] when it comes to making recruitment decisions [Using AI to decide which freelancers to recruit for their joint bid]."}}

\blue{
These findings can help to reveal how current generative AI tools embody technological rationality through context failure that prioritizes standardization over situated knowledge and over-reliance that reshapes human practices toward one-dimensional responses.}


\sout{P4 noted that freelancers often accept AI-generated information without critically evaluating it, treating it as the final answer rather than using it as a starting point for discussion and collaboration with their peers: \textit{"People over-rely on AI [generative AI tools] these days. Like you ask someone, how do you know whether an information the AI gave us is correct? So you see, some people don't even check it [AI generated outputs], they just go, they are not creative..."} }

\sout{These findings reveal how current generative AI tools embody technological rationality through context failure that prioritizes standardization over situated knowledge and over-reliance that reshapes human practices toward one-dimensional responses. Freelancers' critiques of these limitations and their articulation of what human communities provide that AI currently lacks, offer design requirements for collaborative AI that preserve situated knowledge and support rather than undermine human agency in collaborative work.}

\subsection{\blue{Risk of AI-Generated Plagiarism and Loss of Creative Agency}}
\sout{Freelancers in our study raised ethical concerns about generative AI’s implications for originality, plagiarism, creative agency, fairness, and inclusivity. Below, we outline the key ethical challenges they identified in their co-design sessions.} 

\blue{Freelancers often team up to bid for projects and complete the work together \cite{fulker2024cooperation}. In this context, our participants expressed concerns that AI might reproduce content without proper attribution, leading to unintentional plagiarism and making their teams' contributions appear less original. They worried this could damage their reputations and reduce their chances of securing future joint projects, as their teams might be perceived as just relying on unoriginal, AI-generated work. P8 further explained this dynamic: \emph{``of course ChatGPT can do that [help them generate a proposal for a joint job bid], but if you will try a couple of queries [different prompts to the generative AI to create the proposal], you will see same results [same type of text for the proposal the AI generates] and any client can identify at first glance that, okay, this is a generated [generated by AI]. And when you will have a look at the jobs they [clients] post online on Upwork, let's say the client mentions: ``do not write with ChatGPT or use AI to create proposals.'' So right there, we don't get those jobs.''}} 

\textcolor{blue}{This reinforced freelancers’ fears that AI use could be easily detected by clients, undermining their credibility in joint bids.} Freelancers also raised concerns about AI systems compromising their creative agency and professional identity through the unattributed reproduction of content and potential plagiarism that could emerge during collaboration. P24 also expressed concerns about the originality of AI-generated content, noting that AI often produces similar outputs for different users, which could threaten the ability of freelance teams to deliver distinctive creative work: \textit{"The gen AI model 
usually repeats stuff very often. So, like, the originality in the text gets lost."} 

\textcolor{blue}{For freelancers, this repetition could directly threaten their ability to deliver distinctive contributions within a team.} P21 extended this concern to AI-generated images, highlighting risks to creative attribution: \textit{"I get anxious when generating images [with AI tools], for example, I wonder if sometimes the images aren't original. And so what I mean by that is, let's say there are tens of thousands or hundreds of thousands of people who are generating a specific image, and many of them for a specific niche. And I wonder if certain AI tools might create an image that looks exactly the same, or at least very, very similar. Then there's doubt about where it originated. The same thing happens pretty often with text."} 

\textcolor{blue}{This highlights how repetition in AI-generated outputs can threaten freelancers’ sense of creative agency and ownership.} This anxiety also surfaced in collaborative contexts, where participants worried that attribution issues could spill over into their client relationships. P8 expressed concerns that AI tools might reproduce identical content across collaborators, creating potential disputes over creative ownership: \textit{``But how can I be sure that if I write a piece of content [using generative AI] for my client, that ChatGPT is not going to provide that same content to some other person [freelancers in their collaborative project]?’'} \textcolor{blue}{Such duplication risks not only the collaborative process but also the trust freelancers must maintain with their collaborators and clients.}

These concerns reflect freelancers' need to maintain creative agency and professional authenticity in collaborative work. The anxiety about unattributed content reproduction reveals deeper concerns about preserving creative control, professional differentiation, and the ability to offer genuinely distinctive collaborative outputs.
\blue{While participants focused primarily on concerns about AI reproducing content without attribution, these anxieties exist within a broader context where, as we discuss in Section 5.2, their original work may already have been appropriated to train these very AI systems without their knowledge or consent.}

\subsection{Envisioning Principles for Generative AI to Facilitate Freelancer Collaboration}
During the co-design sessions, freelancers \sout{not only identified challenges and ethical concerns related to generative AI tools for collaboration, but also} suggested design principles to make \blue{generative AI} tools more \blue{attuned to freelancers' needs.} 

\subsubsection{Design Principle: Support Collaboration Across Diverse Technical Skill Levels}

\begin{wrapfigure}{r}{0.5\columnwidth}
    \centering
    \vspace{-8pt}
    \includegraphics[width=0.48\columnwidth]{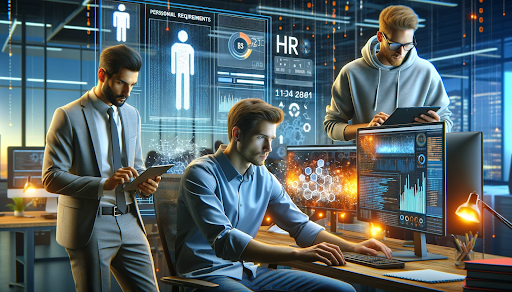}
    \caption{Phase III design probe showing freelancers with diverse skills and background collaborating.}
    \label{fig:Diverse}
    \vspace{-10pt}
\end{wrapfigure}

\blue{Freelancers from different fields often come together to submit joint bids and complete work that requires multiple types of expertise \cite{fulker2024cooperation}. Fig. \ref{fig:Diverse} is a visual probe that depicts freelance collaborators with varying technical skills and backgrounds working together. For our participants, it was thus important that generative AI should support such diverse teams, especially enabling both highly technical and non-technical members to work effectively with the AI. For instance, they envisioned teams where a designer could use natural language prompts while a developer edited code, each interacting with the same AI system through interfaces aligned with their skills and experience. P10 for example, described a design concept for how such collaborative generative AI tools could work:}\textit{"...a community platform [a generative AI enhanced platforms] where there would be both developers [freelancers with deep technical expertise] and other gig members [freelancers with less technical expertise], and it's that one place where all, all of us could unite and put in our opinions and suggestions accordingly [opinions and prompts for the AI]."}

\begin{wrapfigure}{r}{0.5\columnwidth}
    \centering
    \vspace{-8pt}
    \includegraphics[width=0.48\columnwidth]{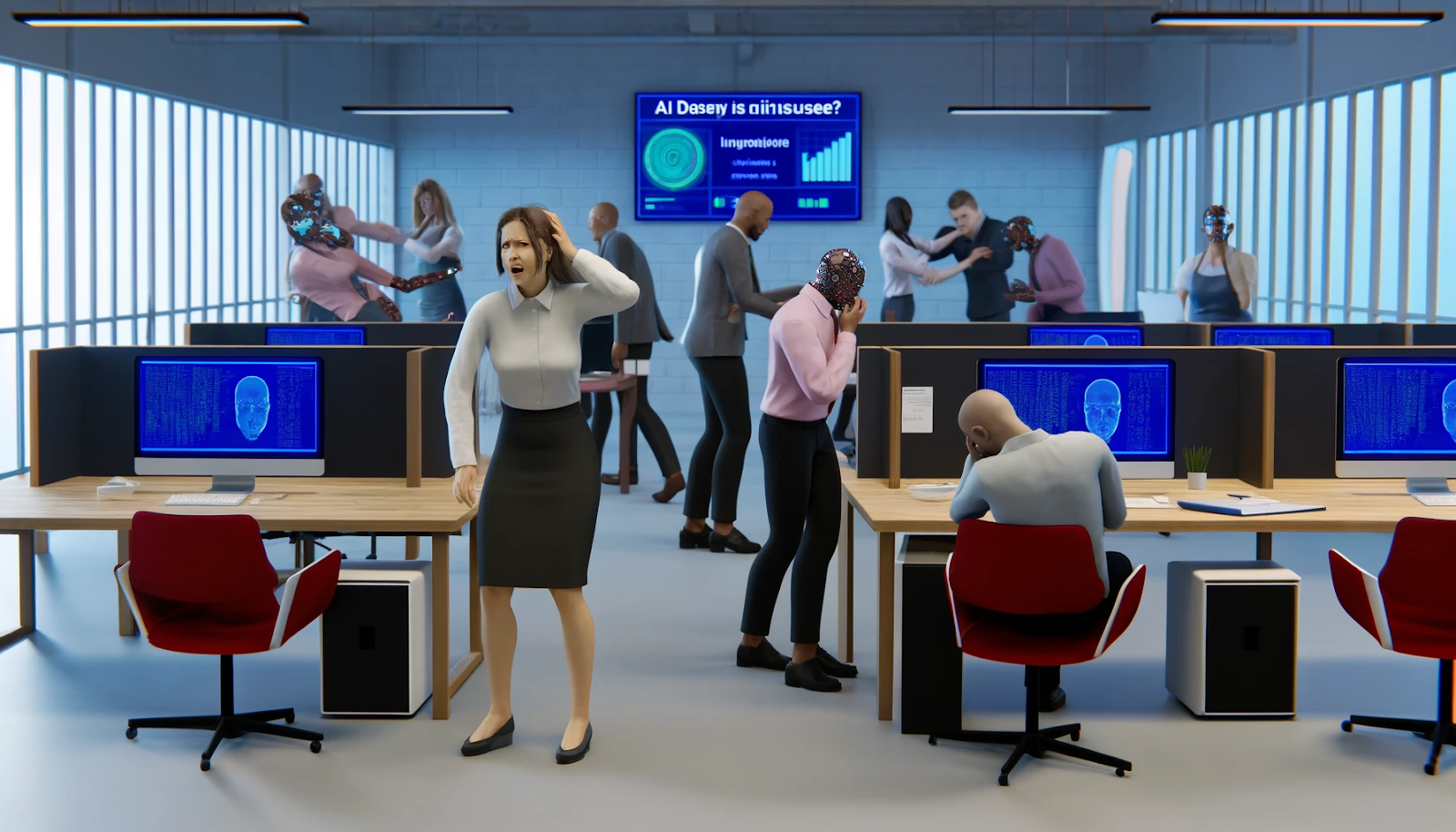}
    \caption{Phase I design probe showing how AI overtakes human oversight when freelancer teams are not diverse and lack technical expertise.}
    \label{fig:AIOvertakes}
    \vspace{-10pt}
\end{wrapfigure}

\blue{This vision of inclusive technical collaboration reflects freelancers' commitment to democratizing participation rather than excluding those with different technical skill levels \cite{DemocratizingTechnologyAndrewFeenberg, ozturkcan2024right}. This design principle ensures that all team members in a collaboration, regardless of technical expertise, can maintain agency and oversight in collaborative processes involving generative AI.} One reason freelancers seemed to value technical diversity in AI-driven collaboration was to prevent over-reliance on AI for decision-making. P13 cautioned against letting AI make technical decisions on its own, emphasizing the importance of human oversight. However, they also pointed out that effective oversight requires freelancers to have the necessary technical expertise to do the decisions on their own: \emph{``So maybe in future, like, if it's project management that we're doing, or giving technical advice, advisory, support, I would not want AI to make the decision on which projects to do or how to schedule the tasks. I would want to be able to be the one to make those decisions...''}. 

This insistence on making crucial decisions themselves highlights how strongly freelancers value maintaining human agency in collaborative work. \blue{Fig. \ref{fig:AIOvertakes} further illustrates this by depicting the risks of teams that lack diversity and technical skills to oversee AI. In this design probe, the AI effectively takes control of the freelancer team because no one had the technical expertise or perspective needed to supervise or challenge its decisions.}

Freelancers also saw collaborative prompt engineering as a way to leverage different skills and improve AI-generated outputs. P22 described how this approach could work in practice: \textit{"We will be researching [...] deciding on some pictures [pictures for their collaboration] giving the instructions on the kind of picture that we want the AI to generate. And then whenever maybe somebody else is not satisfied with that [with the image the AI generated], he can reframe it [reframe the instructions to the AI] in the best way that he knows."} 

This collaborative approach to AI interaction suggests ways to harness diverse expertise while maintaining human control. Freelancers also believed that AI tools should help match people with complementary technical skills, improving team efficiency and project sustainability. P7 proposed using AI-powered matchmaking to connect freelancers with complementary technical skills for co-producing projects:
 \emph{``So AI can be used to tell us or help us to match people: ``Oh, this person [freelancer]''... the AI then is going through their interest and expertise, because AI feeds on vast amount of data. So you can go through this person's interest and expertise, see what this person likes, the kind of jobs they do, and match people. ``Oh, yeah, this person can collaborate more on tech. You and this person have something in common...''} This vision positions AI as a facilitator of human collaboration rather than a replacement for human decision-making.

\subsubsection{Design Principle: AI Should Function as an Auxiliary Team Member, Not the Lead}
\begin{wrapfigure}{r}{0.5\columnwidth}
    \centering
    \vspace{-8pt}
    \includegraphics[width=0.48\columnwidth]{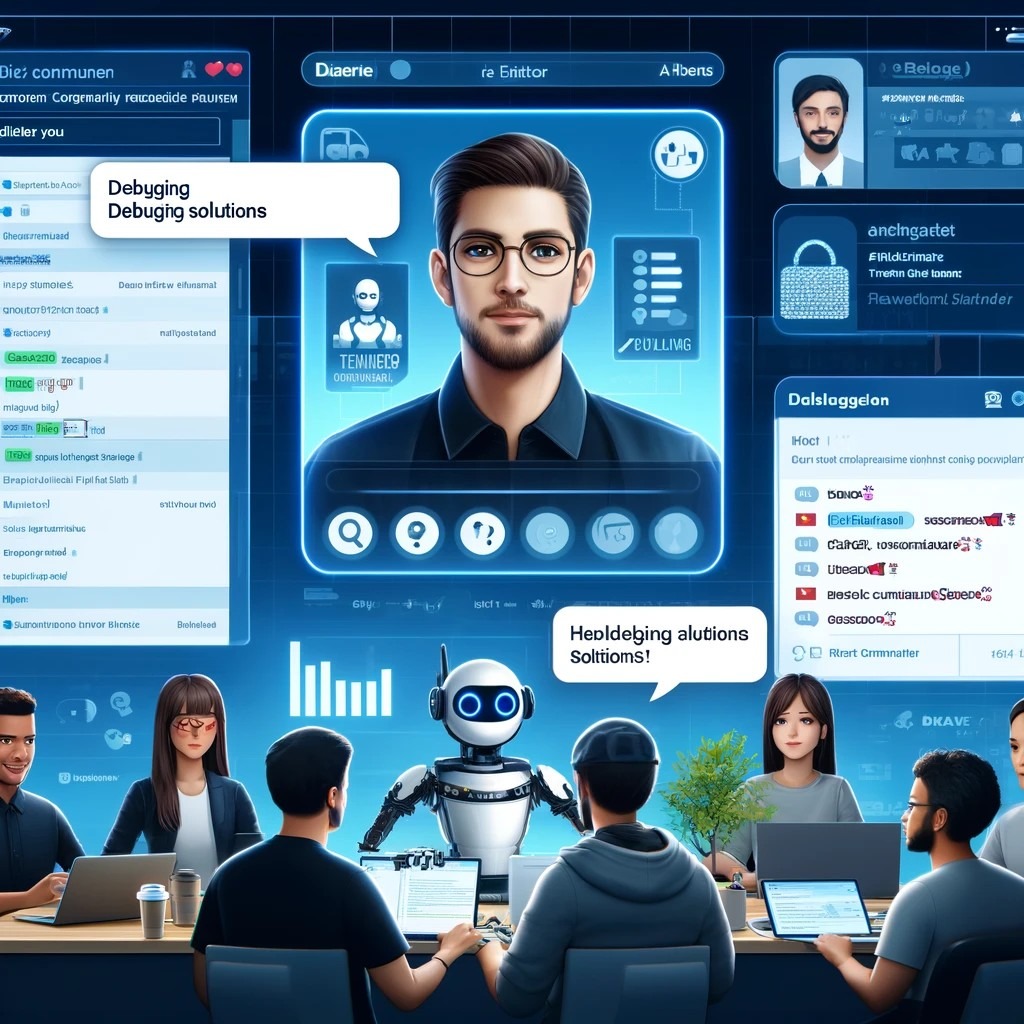}
    \caption{Phase II probe showing participants' vision of auxiliary AI supporting collaboration.}
    \label{fig:auxAI}
    \vspace{-10pt}
\end{wrapfigure}

\blue{Freelancers did not want AI to function as an equal decision-maker in their teams. Instead, they saw it as an auxiliary member that should support their various forms of collaboration, such as creative co-creation, peer learning, and collective action.} \sout{Freelancers rejected the idea of AI as an equal decision-maker in their teams. Instead, they believed it should be an auxiliary team member that supports their collaborations.} Fig. \ref{fig:auxAI} shows a participant-created design probe envisioning AI as a support tool, with a group of freelancers taking the lead, and AI helping them in the background (\textcolor{blue}{see Figure \ref{fig:sharedvisionofAuxAI} in the appendix for more design probe examples of Auxiliary AI}). This vision directly resists automation logic that would position AI as the primary driver of collaborative processes \cite{marcuse2013one}. 

\blue{When imagining how this auxiliary AI should function}, freelancers first recognized that their collaborative projects usually involved individuals with diverse skills \cite{watson2021looking, wood2019good}. Because of this, they viewed generative AI as  auxiliary team member that could help fill any skill gaps within their human collaborations. \sout{For example, P27 envisioned this auxiliary role as follows:}
\textcolor{blue}{For example, P27 described how freelancers with different skills could collaborate to co-create tools, they imagined generative AI acting as an auxiliary team member to support their joint work:}
\textit{``Let's say we have a team of four people [freelancers]. Person 1 is there for tech and recruitment [recruit other freelancers to help co-create the final product]. Person 2 is handling our server side. I'm managing some of the other things. ChatGPT [generative AI tool] acts as a fourth player in our team. While we are discussing, I want ChatGPT 
to give me some numbers [...] I don't expect Person 1 and Person 2 to be good with numbers because they have expertise in different fields, and I think ChatGPT [generative AI tool] can generate those results in the background, supporting our team with real-time data."} 

\sout{Similarly, freelancers noted that AI should assist with advancing the human collaboration.} 
\textcolor{blue}{In addition to helping fill skill gaps, freelancers also imagined AI monitoring collaborative progress and offering ongoing support. For example, participants discussed how AI might assist a team of freelancers developing a social media campaign for a collective action initiative,} and P6 explained: \textit{"The AI can [be] like, providing real time feedback, like monitoring the progress of our initiative or our ideas  and then providing us with a real time feedback that will be great for a team [team of freelancers]."}  

Such a vision of AI as a “fourth player” providing background support reflects the auxiliary AI role that preserves human leadership while leveraging AI capabilities \cite{momose2025human, leeimpactCHI2025}. Freelancers also felt that generative AI, when used as an auxiliary team member, could improve collaboration by helping communication flow more smoothly. \textcolor{blue}{This can be especially important for freelancers, who often work in distributed teams with collaborators around the world \cite{graham2019global} where smooth communication can be difficult and frequent clarification requests can create delays. P13 highlighted how generative AI as an auxiliary team member could help mediate communication between freelancers, particularly when freelancers are collaboratively co-creating tools or documents:} \textit{"In terms of collaborating, I think generative AI can actually make, it can actually provide more clarity in terms of collaboration and also help us [teams of freelancers] save on time we spend in clarifying something. So, let's say if the AI is tracking what you're doing and maybe a colleague or somebody working there needs further clarification, they can just ask the AI and you just continue with your work so at least it can save on time on clarifications."} \textcolor{blue}{This envisions AI as an intermediary that answers routine questions about ongoing work, allowing collaborators to maintain focus on their tasks rather than constantly interrupting each other, such examples show freelancers envisioning AI as enhancing rather than replacing human collaborative capabilities.}

\subsubsection{Design Principle: A Dedicated Human Coordinator Is Key for Maintaining Focus When Working with AI Tools.} 


Freelancers believed that collaborative generative AI tools should include a dedicated human coordinator to guide the 
AI involvement and maintain focus. While AI can assist in managing and streamlining tasks, freelancers expressed concern that AI-generated suggestions could sometimes derail discussions or distract teams from their main objectives, \blue{for example, learning about a topic together, co-creating new artifacts, or jointly bidding on and winning a project proposal.} This emphasis on human coordination reflects freelancers' commitment to preserving human agency in AI-mediated collaboration. P5 described the human coordinator’s role in maintaining team focus: \emph{``...for every project \blue{[AI-mediated freelancer collaboration]}, generally, there would be a project coordinator, right. And part of the assignments or the responsibilities of this project coordinator is to keep everyone focused on the end goal [...] I think having regular meetings or stand ups where we discuss various issues that we might be experiencing in the design process \blue{[co-creation process]} [..] So having such meetings would help us align all of our perspectives [...] So we need to have a project coordinator that, you know, keeps everyone in check, keeps everyone in line [...] It's just really staying focused on what the project really needs.''} Such an emphasis on a coordinator who "keeps everyone focused on the end goal" demonstrates the importance of human oversight in preventing AI from disrupting collaborative processes.

Freelancers found that AI suggestions \blue{in their collaborations}, though helpful at times, could introduce off-topic ideas or shift the conversation unintentionally. \blue{P13 explained how these distractions reinforced the importance of human oversight}: \emph{"...a challenge I've seen is that there's a lot of, there can be distraction. Let's say, for example, you start a discussion, I'm just giving an example. I've gone to a group and I tell them I'm having an issue with ``X''. What should I do? So the first person responds, the second person responds. Then maybe you see later on in the discussion, the AI brings in a new topic, then we tend to get distracted and it ends up being a whole totally different conversation...''} Freelancers' observation about AI bringing in a new topic that leads to distraction highlights the need for human coordination to maintain collaborative focus and intentionality. 

These reflections illustrate the need for a collaboration model, where AI serves an auxiliary role, but human coordination remains essential for keeping \blue{collaborative} projects structured, purposeful, and on track. Freelancers saw the project coordinator not just as a manager, but as a safeguard against AI-induced topic drift, helping to mediate between AI assistance and \blue{freelancers' collaborative} goals.

\subsection{\textcolor{blue}{How DALL·E Helped Surface Participants’ Visions for Collaborative Generative AI}}

\textcolor{blue}{Freelancers in our study came from diverse professional and educational backgrounds  (See Appendix Table \ref{tab:participants} for details), as a result, many lacked formal design training or established ways of visually communicating their ideas. However, because our co-design sessions required participants to speculate about futures of generative AI that do not yet exist, we needed a method that could help freelancers express abstract, emerging, or partially formed ideas. To support this, we used DALL·E as a co-design probe.} \textcolor{blue}{DALL·E enabled forms of speculative ideation that would have been difficult to elicit through traditional methods such as design workbooks, static visual probes, or sketching exercises \cite{auger2013speculative}. Traditional probes typically require participants to translate ideas into visual formats using their own design skills \cite{bannon2012design, williams2002self}. But, this can limit creative expression among individuals without design expertise \cite{williams2002self, harrington2021eliciting}. Likewise, non-AI-generated design probes are usually pre-constructed and can only capture futures that the research team has already anticipated \cite{auger2013speculative, designprobeswallace2013making}. In contrast, DALL·E dynamically generated new visual interpretations of participants’ prompts in real time, introducing unexpected imagery that participants could react to, refine, or reinterpret. These moments of surprise or misalignment surfaced latent assumptions, concerns, and aspirations that would likely not have emerged through conversation or static probes alone.} \textcolor{blue}{For example, we asked freelancers to use DALL·E to imagine future collaborations with AI. Participants often began with vague or incomplete visions, but the AI-generated images helped them externalize ideas they struggled to verbalize. P27 described how DALL·E enabled them to ground and articulate their thinking:
\emph{“...I wrote this prompt [prompt given to DALL·E during our study]: ‘give me an image and the image should consist of: we are in the future and AI is coding itself and it does not require any developer or human interference and AI is confident.’ So, I think sometimes in our prompts , sometimes we even don't know like what we want, right? I just gave him [DALL·E] a thought and I don't know how the future will look like in 40 years or 20 years, right? [...] Sometimes we are just grasping or you know sometimes just brainstorming along with AI, like what can be the possibilities, and maybe AI can generate some unique image and we may not have thought about that particular thing.”}}

\textcolor{blue}{These AI-generated reinterpretations acted as provocations that pushed participants to clarify what aspects of future AI-supported collaboration they found appealing, concerning, or technically plausible. This type of iterative sensemaking would have  likely been difficult to produce through sketches or verbal discussions alone \cite{williams2002self}.}

\textcolor{blue}{DALL·E also enabled dynamic, iterative refinement. Participants issued follow-up prompts to adjust or improve the generated images, gradually sharpening their design requirements. P22 described this process:
\emph{“And looking for a particular thing [their vision of future AI-supported collaboration] in my mind and getting it generated [getting the image generated with DALL-E]... it's not exactly what I expected, but it is towards that what I had in mind. 
I think with more... maybe giving more details or re-framing the question [prompt given to DALL-E], I think I can get close to what I need... I've gotten the images that are within the bracket of what I needed based on the question.”}}

\textcolor{blue}{Through this iterative prompting process, participants not only refined the images but also discovered design criteria and collaborative workflow needs they had not initially articulated. This reflective, step-by-step clarification is a unique affordance of AI-generated imagery compared to non-AI probes.}

\textcolor{blue}{Importantly, DALL·E’s text-based interface lowered barriers to visual ideation. All freelancers in our study, including those without design backgrounds, were able to produce meaningful visual concepts. P21 noted that DALL·E helped them generate a visual probe directly connected to their professional context:
\emph{“I'm a freelancer on Upwork, and I certainly do collaborative work in digital marketing, creating images and text for social media and for advertising... I asked the AI tool [DALL-E in our co-design sessions] to create an image focused on that area. And so I think it came out pretty cool. Certainly, it [image of a person generated by DALL-E] looks like she's working with freelancers and creating images and generative text through generative AI tools for the purpose of gig freelancer work.”}}

\textcolor{blue}{This accessibility allowed participants with widely differing expertise to contribute equally to the speculative design process, something traditional sketch-based methods likely would not have easily supported \cite{bressa2019sketching}.}

\textcolor{blue}{Finally, DALL·E-generated images appeared to serve as boundary objects that sparked dialogue, negotiation, and collaborative refinement among participants \cite{zamenopoulos2021types}. These images appeared to provide concrete, shareable reference points that freelancers could collectively critique and reshape. P22 illustrated this dynamic:
\emph{“You can collaborate with people [freelancers in our co-design sessions] in different regions [...] and everyone of us is giving ideas [...] maybe we [group of freelancers in our co-design sessions] want to see some pictures. At some point, we will be writing, we're giving the instructions on the kind of picture that we want the AI to generate. And then whenever maybe somebody else [one of the freelancers in our co-design sessions] is not satisfied with that [with the image the AI generated], he can re-frame it in the best way that he knows. At the end of the day, we 
will be coming up with very nice designs that we will have generated from working together.”}}

DALL·E's collaborative refinement process helped transform abstract ideas into concrete design features, surfacing insights about AI-supported collaboration that traditional probes cannot easily capture. Examples of this iterative refinement process and participant-generated visions across co-design sessions are presented in Appendix \ref{appeni}.

\begin{figure*}[htbp]
  \centering
  \includegraphics[width=0.75\linewidth]{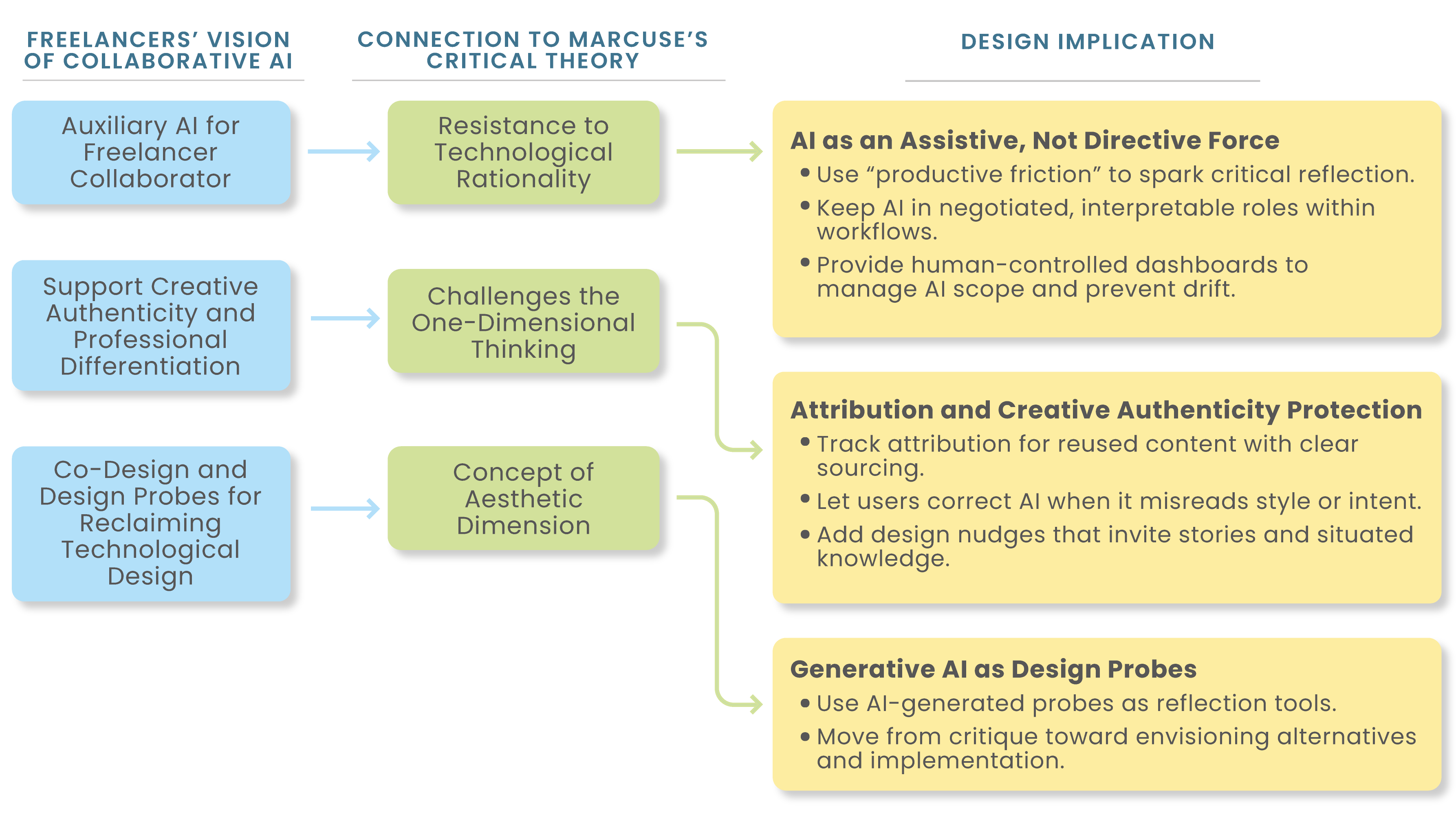}
  \caption{Freelancers’ vision of Collaborative AI, Connection to Marcuse, and Design implementations.}
  \Description{Flow diagram illustrating how freelancers’ visions of collaborative AI (left) connect to Marcuse’s critical theory (center) and lead to design implications (right). On the left are three visions: auxiliary AI for freelancers, support for creative authenticity, and co-design with design probes. These connect to Marcuse’s concepts in the center: resisting technological rationality, challenging one-dimensional thinking, and emphasizing the aesthetic dimension. On the right, the diagram shows design implications, including positioning AI as an assistive force, protecting creative authenticity, and using generative AI as design probes.}
  \label{designGuidelines}
\end{figure*}

\section{Discussion}
Our co-design study reveals tensions between how freelancers currently experience generative AI in collaborative work and how they envisioned these tools should be designed. Drawing on Marcuse’s theory of technological rationality \cite{marcuse2013one}, we argue that these tensions reflect a deeper conflict between efficiency-driven AI design and the participatory, situated knowledge that is essential for meaningful collaboration. In the following, we discuss these tensions in detail. Figure \ref{designGuidelines} connects freelancers' collaborative AI vision to Marcuse's theory and design implications.

\subsection{Auxiliary AI: A Vision for Future Generative AI Tools for Freelancer Collaboration}
Recent research shows that over-reliance on AI can weaken individual decision-making in traditional organizations \cite{leeimpactCHI2025}. Freelancers in our study extended this concern to collaboration, warning that AI dependency could undermine the shared focus needed for joint work. 
As freelancers collaborate without institutional support or managers who can realign goals \cite{jarrahi2021flexible,sutherland2017gig}, even small shifts in attention caused by AI can have amplified consequences. Participants described anxiety about “topic drift,” where AI suggestions were technically relevant but misaligned with the team’s direction, quietly redirecting collaboration in ways that were difficult to notice or correct.

Working entirely at a distance, freelancers rely on the digital workspace as their only shared environment \cite{jarrahi2020platformic,sutherland2018sharing}. If AI shapes this space too strongly, they feared losing alignment and coordination.
These concerns might help explain why freelancers envisioned generative AI as an auxiliary supporter that strengthens human-led teamwork and keeps collaborators grounded in intentional human dialogue \cite{dignum2001creating}.

This vision contrasts with many current collaborative Generative AI tools, which often organize teamwork around the AI interface instead of around people \cite{han2024teams,shaer2024ai,wang2025aideation}. Platforms such as ChatGPT Teams \cite{openai_chatgpt_team_2024}, Claude Projects \cite{anthropic_projects_2024}, and GitHub Copilot Workspace \cite{github_copilot_workspace_2024} offer powerful shared features, but they still mediate collaboration through AI-driven workflows. Prior research shows that most generative AI systems frame collaboration as joint interaction with the AI, rather than as interpersonal negotiation or human-to-human sense-making \cite{seymour2024speculating, shneiderman2022human, johnson2025exploring}. For freelancers whose income depends on staying aligned with collaborators and clients \cite{fulker2024cooperation}, this AI-centered model can increase the risk of drifting out of sync or missing commitments in ways that may jeopardize contracts \cite{spektor2023designing, jarrahi2020platformic}.

To address the limitations of current generative AI tools for collaboration, freelancers in our study imagined future systems not as the center of teamwork, but as auxiliary collaborators that quietly support human-to-human interaction. In their vision, AI should step back from creative and strategic decision-making. By keeping AI in a background role and designing interfaces that protect time for human-to-human discussion, freelancers believed this arrangement would strengthen human creative agency, allowing AI to support the work without competing for control of the project.

Freelancers also wanted auxiliary AI systems that could adapt to varied skill levels, giving AI experts rich ways to engage with the system while keeping interfaces accessible for non-experts. They argued that teams should be able to edit, reconfigure, and extend the AI systems they use rather than having to work around closed, proprietary tools \cite{bommasani2021opportunities, chang2024survey}. Most current generative AI tools are designed for narrow ideal users \cite{zamfirescu2023johnny, sambasivan2021everyone, leeimpactCHI2025}, provide limited support for multiple roles, and rarely allow teams to modify models or workflows \cite{wellner2020feminist,holstein2019improving}. As a result, the kind of deep democratization Feenberg describes, where users can embed their own values and interventions into technical structures \cite{DemocratizingTechnologyAndrewFeenberg}, is largely missing from today’s AI collaborative systems.

Freelancers’ proposals also resonated with right-to-repair movements \cite{ozturkcan2024right} and rejected one-size-fits-all AI systems \cite{onesizefitsallCHI23}. They sought to foreground equity in how people interact and work with AI by addressing disparities in skill, authority, and voice within teams \cite{ehsan2020human, ehsan2021expanding}. They emphasized that AI should support, rather than override, the social dynamics of collaboration. This is likely also why they wanted non-AI experts to be able to collaborate easily with the AI; otherwise, the AI risks creating a new power dynamic where only technical experts lead the collaboration. Freelancers' call for modifiable and reconfigurable systems reflects a vision of deeply democratized generative AI that current tools do not yet support.

\subsubsection{Design Implications:} Freelancers in our study did not want to simply follow AI-generated suggestions. They wanted to engage critically with 
AI suggestion and retain human agency in collaboration. In this context, we argue that auxiliary AI systems should intentionally introduce “productive friction,” where automated responses are slowed or made less seamless to encourage reflection, discussion, and critical engagement among freelance collaborators \cite{xu2025productive}. Rather than streamlining collaboration until shared ideas are scripted by AI or group experimentation is fully automated \cite{horton2023large, manning2024automated}, auxiliary AI can help preserve intentional pauses, negotiation, and interpretation within workflows \cite{AgencyinHumanAIcollab}. These frictions are not flaws but design features that resist seamless automation of collective decision-making \cite{huang2023generative}.

Implementing productive friction requires specific interface patterns. Human coordinator dashboards should let designated team members monitor AI involvement, filter suggestions for relevance, and maintain focus, directly addressing the “topic drift” freelancers identified. Collaborative prompt engineering interfaces should support teams in refining AI instructions together, allowing members with different technical skills to contribute through visual templates and iterative improvement. Interfaces should also foreground AI’s auxiliary role by positioning AI contributions as supplementary to human dialogue rather than central decision-makers. This includes explicit human-controlled activation of AI suggestions and visual hierarchies that prioritize human input over AI-generated content, embodying freelancers’ vision of AI as a background supporter rather than a director.

\subsection{Originality and Creative Attribution in Generative AI–Mediated Freelancer Collaboration}
Freelancers raised concerns that generative AI could erode their creative agency and professional identity, especially in activities such as creative co-production and joint bidding. They worried that AI might reproduce their ideas without attribution or misinterpret their stylistic intentions, weakening the authenticity and distinctiveness of their shared work and ultimately jeopardizing their bids and income.

These concerns are likely tied to how generative AI assigns credit and interprets creative intent, which is shaped by the stochastic nature of these systems \cite{he2025contributions}. Generative models rely on probabilistic processes \cite{holtzman2019curious, hao2023optimizing}, so a prompt such as “a cat on a mat” does not specify a single outcome but triggers random decisions about lighting, texture, syntax, and tone in ways the user cannot fully control \cite{hao2023optimizing}. This weakens the causal link between intent and output and allows models to produce strikingly similar results across different users \cite{zamfirescu2023johnny}. Freelancers consequently feared that their work could become indistinguishable from others, making it harder to demonstrate the unique value of their collaboration. These worries are intensified by broader limitations in how generative AI interprets style, assigns credit, and supports collaborative originality \cite{he2025exploring}, as well as persistent human biases in judging AI versus human contributions \cite{he2025contributions}. When these AI systems are widely embedded into collaborative workflows, they risk creating a “tragedy of the generative commons” \cite{huang2023generative, ostrom1990governing}, where widespread use of tools that reproduce unattributed content and blur stylistic boundaries can diminish the creative integrity, richness, and originality of collective work \cite{kadoma2024role, jakesch2023co}.

Freelancers’ concerns also reflect a broader structural reality. Generative models are trained on vast amounts of human-created content, often without creators’ awareness or consent \cite{dornis2025generative}, resulting in large-scale appropriation that has already occurred \cite{chesterman2025good,chiba2025tackling}. Large technology companies capture value from collective human creativity \cite{MacCarthy2025,kyi2025governancegenerativeaicreative}, while individual creators, e.g., freelancers with limited institutional power, usually have few mechanisms to protect their work or shape how it is used \cite{gray2019ghost,tubaro2020trainer}. Even as freelancers propose alternative designs, they operate in a landscape where the terms of appropriation were set long before they entered the collaboration \cite{do2024designing, dontcheva2014combining}.

These dynamics affect not only labor but also the integrity of creative outputs. \citet{agarwal2025ai, agarwal2025fluent} show that AI systems can erase cultural nuance and nudge users, especially those from non-Western backgrounds, toward Western stylistic conventions, often without users realizing that their intended style has been misinterpreted. This can be particularly damaging for freelancers whose competitive advantage sometimes depends on offering diverse cultural perspectives and distinctive stylistic approaches \cite{bogenhold2016independent}. When AI tools misinterpret or override these situated creative practices, freelancers lose key sources of professional differentiation that enable them to serve specialized market niches with authentic, contextual expertise \cite{bogenhold2016independent, haraway2013situated}. This likely explains why freelancers feared that losing their creative voice would undermine both creative co-creation and joint bidding. If their work begins to resemble generic AI output, clients may no longer recognize their unique perspectives and skills, making them less competitive and more likely to lose bids.

Overall, this empirical evidence supports our participants’ concern that generative AI can undermine the originality and culturally situated perspectives that give their collaborations value \cite{agarwal2025ai}. This erosion of creative agency reflects Marcuse’s notion of “one-dimensional thinking,” where technological systems compress diverse creative possibilities into standardized, reproducible outputs and present appropriation and standardization as both efficient and inevitable \cite{marcuse2013one}. Feminist theorists such as \citet{haraway2013situated} show how these processes erase situated knowledge, while critical computing scholars document how AI universalizes dominant cultural patterns and sidelines contextual approaches \cite{septiandri2023weird, bardzell2018critical, agarwal2025ai}. Together, these perspectives clarify that freelancers’ resistance to AI’s impact on their creative agency is part of a broader struggle to protect authentic, contextually grounded collaborative creativity.

\subsubsection{Design Implications:}
CSCW has examined how teams negotiate credit through commit logs, meeting notes, and shared norms about who did what \cite{dabbish2012social}. Generative AI disrupts these arrangements \cite{he2025contributions, vaccaro2024combinations}. It introduces a free-rider problem \cite{hampton1987free}, since heavy reliance on AI can blur the boundary between human effort and automated output \citet{shu2024towards}, and it creates social friction when teams lack norms for recognizing AI’s role \cite{he2025exploring}. In response, new frameworks and toolkits study how humans and AI work together and how their contributions should be coordinated and credited \cite{AIAttributionToolkit, bagehorn2025ai}. For example, approaches such as PeopleJoin and Collaborative Generative AI Agents shift attention from individual authorship to collective efficacy \cite{johnson2025exploring, jhamtani2025lm}, treating AI as a shared “thought partner” and emphasizing process-based credit such as who coordinated and leveraged AI effectively \cite{kim2021collaborative, tankelevitch2025understanding}. IBM’s AI Attribution Toolkit offers a complementary taxonomy-based approach \cite{AIAttributionToolkit, bagehorn2025ai}, classifying contributions by initiative, volume, and nature and attaching machine-readable metadata so downstream users can see how AI participated and calibrate their trust.

Extending these ideas, we propose AI tools that protect attribution and preserve creative expression in collaborative workflows. Such systems would provide clear credit for reused content, make stylistic interpretations transparent, and flag moments when creative approaches are misrepresented or insufficiently acknowledged. This aligns with reflective design and critical making, which encourage users to question and reshape technology rather than accept its outputs uncritically \cite{sengers2005reflective, ratto2011critical}. Drawing on Marcuse’s aesthetic dimension \cite{marcuse2014aesthetic, marcuse2013one, bardzell2018critical}, we argue that future AI tools for freelancers should embed authentic creative expression at their core. Concretely, they could support transparent attribution tracking, user validation and correction of stylistic interpretations, and customization of outputs to reflect local knowledge, community aesthetics, and individual priorities. This is especially important given evidence that AI systems often erase cultural nuance and nudge creators toward Western stylistic norms, undermining the aesthetic and economic value of non-Western distinctiveness \cite{agarwal2025ai, agarwal2025fluent}. Introducing productive friction can help users notice when AI suggestions override cultural specificity and preserve the situated knowledge that makes their contributions valuable.

\subsection{Using Co-Design and Generative Design Probes for Reclaiming Technological Design}
A challenge in co-design is that people often struggle to imagine possibilities beyond the tools and artifacts they already know \cite{jansson1991design}. However with DALL·E, freelancers were able to move beyond critique and begin articulating concrete design alternatives. This showed how generative AI systems can function as co-design probes that help freelancers collectively envision alternative implementations of the very technologies they are critiquing. In line with probes being “sites and tools for reflection” \cite{designprobeswallace2013making} and “supports for overcoming self-censorship” \cite{williams2002self}, we observed that the generative AI images participants produced did indeed act as probes, helping freelancers express aspirations that might not surface in purely verbal discussion. Participants described these probes as useful for expanding their imagination and helping them articulate alternative realities for how generative AI might support collaboration.

Importantly, enabling freelancers to envision alternative futures was not simply a matter of giving them generative AI tools and asking them to create images. Building on prior work that uses generative AI to support  designers’ creativity and speculation \cite{cai2023designaid, wang2025exploring, lc2023speculative}, we engaged freelancers in a Future Workshop \cite{kensing2020generating} to collectively imagine collaborative generative AI. DALL·E offered several advantages as a probe in these workshops. Unlike fixed probes commonly used in other design methods such as pre-made cultural probe kits, speculative artifacts, or static image sets \cite{brandt2006designing, auger2013speculative, harrington2021eliciting, zamenopoulos2021types}, DALL·E generated probes in real time. This dynamic generation allowed participants to immediately iterate on ideas, react to emerging imagery, and collaboratively shape the probes themselves. DALL·E's text-based interface also lowered barriers for participants without design training, enabling freelancers to participate more fully and articulate their visions \cite{bannon2012design}. As our results showed, iterative prompting surfaced unexpected directions that participants then refined, illustrating how generative AI can broaden speculative exploration in co-design.

Unlike prior work where software teams or other experts co-prototype AI applications \cite{subramonyam2025prototyping}, our approach enabled end-users themselves (many of whom are not AI experts), to direct AI design for their own collaborative needs. This helped position freelancers as co-creators rather than passive recipients \cite{costanza2021design}. Shifting from asking “how will AI transform collaboration?” to “how should AI support the collaboration we (freelancers) want?” aligns with critical design efforts to amplify underserved voices \cite{menendez2017fostering, sciannamblo2021caring, ChallangesinCriticalDesign}. It also demonstrates how AI-enabled probes, combined with structured reflection, can foster a critical consciousness that resists efficiency-driven human–AI interaction and instead foregrounds the forms of human–human collaboration freelancers value.

\subsubsection{Design Implication:}
Our findings suggest several design directions for AI-enabled co-design tools. First, tools should offer visual, probe-based editors where people iteratively adjust AI-generated images or interface sketches, so that requirements are captured through direct manipulation rather than technical specification \cite{williams2002self}. Interfaces could also use prompt iteration to intentionally surface alternative design directions and support exploration of the design space. Second, co-design platforms should support hybrid synchronous and asynchronous participation by storing prompts, probes, and annotations in shared workspaces that participants can revisit, remix, and comment on over time, accommodating time poverty and distributed work \cite{vickery1977time}. Third, systems should treat current AI tools as design materials, providing workflows for critiquing existing tools, generating alternative variants, and comparing them side by side as part of structured reflection. Finally, integrating rapid text-to-UI 
generative AI tools (for example, Vercel v0 or Claude) into co-design processes would let people turn speculative concepts into interactive mockups, enabling more embodied critique and iteration on alternatives to efficiency-driven automation \cite{v0dev, claude}.

\subsubsection*{Limitations and Future Work}  While our study offers valuable insights, it did not include freelancers who were strongly skeptical of AI, and our sample may not reflect the full diversity of freelance skills, specialties, and platforms. Future work should engage a broader range of freelancers, including skeptics, to capture more varied perspectives. \textcolor{blue}{Our use of DALL·E as a co-design probe also created methodological tension, as participants critiqued generative AI while using a tool that reproduces similar issues of appropriation and bias. However, this direct engagement enabled more grounded critiques based on hands-on experience rather than abstract concerns.} In addition, we focused on workers’ perceptions rather than the actual collaborative performance of AI tools. Future studies could include controlled comparisons of collaborations with and without AI assistance, as well as longitudinal research that tracks how freelancers’ relationships with collaborative AI evolve over time and shape their work practices and professional identities.

\section{Conclusion}
\blue{In this paper, we conducted co-design sessions with 27 freelancers to examine the limitations of current generative AI tools for collaboration and how freelancers imagine their future. Rather than seeking more efficient automation, participants envisioned “auxiliary AI” systems that remain in a supportive role, handling coordination, exploration, and documentation while leaving freelancers in control of creative and strategic decisions. They wanted systems that foster productive friction, 
preserve clear attribution, and protect diverse contributions. This vision challenges one-dimensional, efficiency-driven AI and contrasts with tools that center interaction on the AI interface, since freelancers wanted AI to strengthen human–human collaboration.}

\sout{This paper addresses how current generative AI tools embody technological rationality by prioritizing individual productivity over collective engagement. Through co-design sessions with 27 freelancers, we used DALL·E as a design probe tool to envision alternative collaborative AI systems \textcolor{blue}{for freelancers}.
Participants identified ethical challenges and rejected AI-led workflows during their collaboration, instead proposing "auxiliary AI" systems that support human-human collaboration through creative authenticity protection during collaboration. Their visions resist efficiency-driven automation by preserving human agency and situated knowledge in collaborative decision-making.
Drawing on Marcuse's critical theory, we demonstrate how freelancers actively challenge one-dimensional thinking by reimagining AI as background supporter rather than replacement for human collaboration. We conclude with interface design principles for collaborative AI \textcolor{blue}{for freelancers} that enables rich, human-centered teamwork instead of reinforcing technological rationality's isolating logic.}

{\bf ACKNOWLEDGMENTS.} Special thanks to all the anonymous reviewers who helped us to strengthen the paper as well as the freelancers who participated in the study. This work was partially supported by NSF grants 2339443 and  2403252.

\bibliographystyle{ACM-Reference-Format}
\bibliography{Cite}

\clearpage

\appendix

\textcolor{blue}{\section{Appendix}}
\subsection{Forms of Collaboration in Freelance Work}
\label{appendix:collab-types}

\subsubsection{\blue{Freelancer Collaboration: Peer Learning}}
\blue{Freelancers typically lack the built-in learning channels of traditional employment \cite{wilkins2022gigified}. To compensate, they deliberately build peer-learning networks \cite{gray2019ghost, hayes2018networked}, using social media groups, chat channels, and community-organized workshops for training and informal education \cite{margaryan2016understanding,drechsler2025systematic,dontcheva2014combining}. These networks substitute for organizational scaffolding and support freelancers’ upskilling \cite{wilkins2022gigified,chiang2018crowd}.}

\subsubsection{\blue{Freelancer Collaboration: Subcontracting}} \blue{Another form of freelancer collaboration is subcontracting \cite{morris2017subcontracting}. Freelancers subcontract parts of a project when they lack specific skills, face tight deadlines, or handle multi-component work \cite{kittur2013future}. One freelancer serves as the primary worker responsible for final delivery, while delegating specific pieces to secondary workers \cite{morris2017subcontracting, pettas2024platform}. Subcontracting can support skill development, help newcomers build reputation, and better match tasks to workers’ strengths \cite{suzuki2016atelier, gray2016crowd, hui2019distributed}. It may involve real-time help, decomposing tasks, or refining instructions to make work clearer and more efficient \cite{dow2012shepherding,rechkemmer2022understanding}. When done fairly, subcontracting allows freelancers to support each other, share opportunities, and produce higher quality work together \cite{morris2017subcontracting, 10.1145/3359227, wallace2025towards}.}

\subsubsection{\blue{Freelancer Collaboration: Joint Bidding}}
\blue{On most online labor platforms, freelancers bid for projects with a short pitch, evidence of prior work, a price, and a timeline \cite{hsieh2022little, kim2023online}. Because competition is high, freelancers often form ad hoc teams or join crowdfarms, loose collectives that pool skills to submit joint proposals for end-to-end work \cite{wang2020crowdsourcing, wang2021examination}. Typically, a lead freelancer coordinates client communication and assembles the proposal, while teammates contribute scoped sections, cost estimates, and portfolio examples in their specialties \cite{wang2021examination}. Recent work finds that many freelancers are willing to collaborate on team-based bids when projects require multiple skills \cite{fulker2024cooperation}.}

\subsubsection{\blue{Freelancer Collaboration: Creative Co-Production }} \blue{Freelancers increasingly team up for creative work in design, writing, marketing, and media \cite{oppenlaender2020creativity}. They form informal peer networks and ad hoc partnerships to trade critique, divide specialized roles, and keep final products coherent \cite{strunk2023building, kinder2019gig}. This collaboration involves substantial invisible work to align expectations, set shared goals, and maintain accountability, resembling the effort of running a small creative studio \cite{toxtli2021quantifying, hulikal2022collaboration,gray2019ghost}. HCI research has explored systems that support these practices. Flash Teams and Flash Organizations show that short-lived teams can perform well when platforms provide clear roles, scripted workflows, and structured handoffs \cite{valentine2017flash}. Systems like Mechanical Novel demonstrate that distributed contributors can create cohesive artifacts when collaboration includes reflect–revise cycles, shared critique structures, and intermediate prototypes \cite{kim2017mechanical}, while tools, e.g., CrowdCrit \cite{luther2014crowdcrit}, highlight feedback as a central site where freelancers negotiate tone and narrative direction.}

\subsection{\blue{Freelancer Collaboration: Tools for Cooperation \& Collective Action}}
\blue{Beyond creative work, several technologies explicitly support freelancer collaboration and collective action. Reputation and alert systems let workers rate requesters and flag exploitative jobs, helping peers avoid risk \cite{savage2020becoming, saito2019turkscanner, toxtli2020reputation}. Collective-action tools enable petition campaigns, wage analysis, and pay-transparency efforts that coordinate dispersed workers \cite{hsieh2025gig2gether, imteyaz2024gigsense}. Platform-specific mobilization tools have also supported actions such as the Christmas letter to Amazon’s CEO demanding better conditions \cite{gray2019ghost, cherry2019global}. Together, these tools reduce uncertainty, strengthen bargaining power, and counter platform asymmetries \cite{wood2019good,poderi2017trebor}.}

\label{appendix:appen}

\section{Participant Demographic Table}

\begin{table}[htbp]
\centering
\caption{Participant Demographics in our study.}   
\label{tab:participants}
\scriptsize
\setlength{\tabcolsep}{2pt}
\begin{tabular}{p{0.5cm} p{0.5cm} p{0.5cm} p{1.8cm} p{1.8cm}}
\toprule
ID & Age & Gen. & Platforms & Expertise \\
\midrule
P1 & 35 & F & Upwork, Fiverr & Virtual Assistant \\
P2 & 34 & M & Upwork & Data Science \\
P3 & 25 & M & Upwork, Toloka, Oneforma & Translation, Data Analysis \\
P4 & 21 & F & Upwork, LinkedIn & Social Media, Prompt Eng. \\
P5 & 26 & M & Upwork, Toloka, Appen & VA, Research, Design \\
P6 & 27 & F & Fiverr, Upwork & Data Analyst \\
P7 & 30 & F & Upwork, Freelancer & Product Management \\
P8 & 30 & F & Upwork, Freelancer & Full Stack Dev, SEO \\
P9 & 25 & M & Upwork, Freelancer & Data Analyst \\
P10 & 22 & F & LinkedIn, Fiverr, Upwork & Content Writing, SEO \\
P11 & 23 & M & Fiverr & Digital Art, ML \\
P12 & 31 & F & Upwork & Full Stack Dev \\
P13 & 35 & F & Upwork, Braintrust, Fiverr & Project Mgmt, Marketing \\
P14 & 30 & M & Upwork & Legal, Content Writing \\
P15 & 32 & M & Upwork, Fiverr & Product Marketing \\
P16 & 34 & M & Fiverr, Upwork, Freelancer & Translation, Editing \\
P17 & 22 & F & Upwork, Fiverr & UI/UX Design \\
P18 & 27 & M & Upwork & Video Editing \\
P19 & 29 & F & Upwork, Freelancer & Data Entry \\
P20 & 31 & F & Upwork, Freelancer & Recruitment, Coaching \\
P21 & 39 & M & Upwork, Doordash & Graphic Design \\
P22 & 25 & M & Course Hero, Fiverr & DevOps, AWS \\
P23 & 38 & M & Upwork, Fiverr & HR, Digital Marketing \\
P24 & 24 & M & Upwork, Fiverr & SEO, Graphic Design \\
P25 & 18 & M & Upwork, Fiverr & UI/UX, Technical Writing \\
P26 & 39 & M & Upwork & HR, Career Coach \\
P27 & 26 & M & LinkedIn, Upwork & Software Dev \\
\bottomrule
\end{tabular}
\end{table}

\section{Participant-Generated Visions of Collaborative AI Through DALL-E}
\label{appeni}
\subsection{Iterative Refinement Within Co-Design Session}

\begin{figure*}[htbp]
    \centering
    \includegraphics[width=0.8\linewidth]{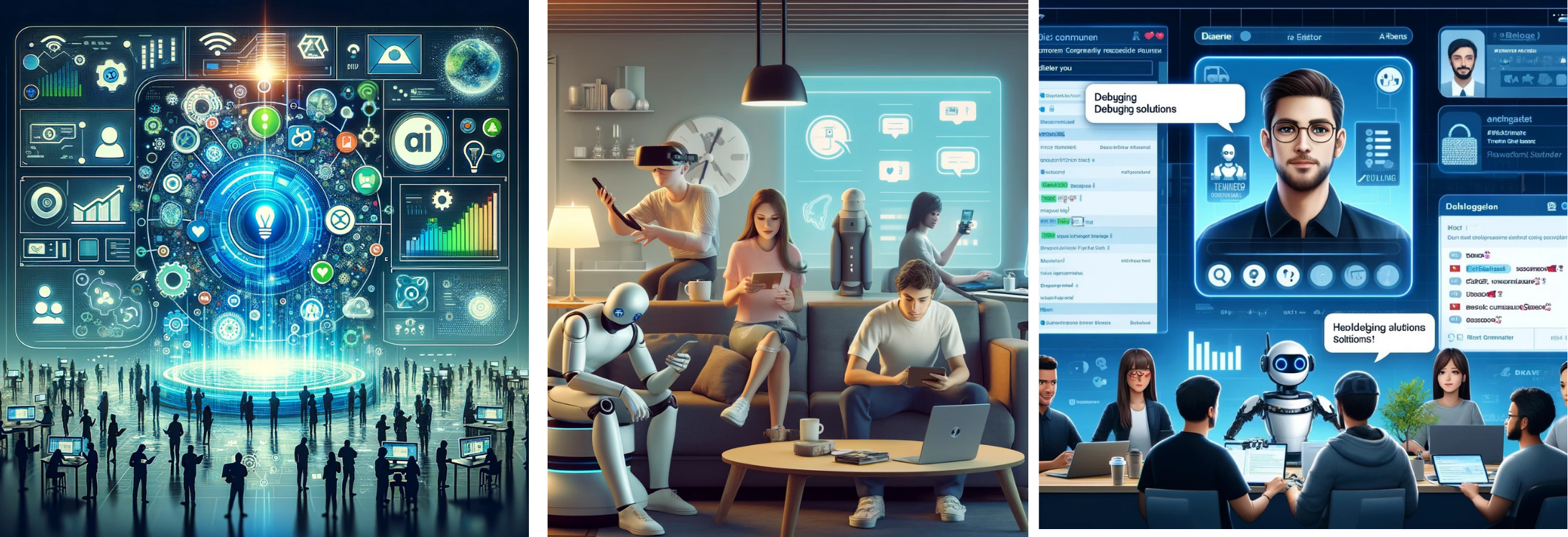}
    \caption{Iterative DALL-E outputs from Co-design (Group 3). Left (Iteration 1): Abstract technological infrastructure with peripheral human presence. Center (Iteration 2): AI robot as equal collaborator. Right (Iteration 3): Final "auxiliary AI" vision with humans foregrounded as decision-makers and AI providing background support.}
    \Description{Three DALL-E generated images showing iterative refinement of collaborative AI vision. Left panel shows abstract blue technological infrastructure with circuit patterns and minimal human presence. Center panel depicts a humanoid robot sitting at a table as an equal participant alongside human workers. Right panel shows a group of diverse professionals collaborating in the foreground with holographic AI interfaces and data displays visible in the background, representing AI as supportive infrastructure rather than central actor.}
    \label{fig:improvedd-probe}
\end{figure*}

Figure \ref{fig:improvedd-probe} illustrates how participants from Group 3 collaboratively and iteratively refined their use of DALL-E during the co-design session. Through this joint refinement process, they began specifying spatial relationships and role hierarchies in their visual probes, moving from more abstract, technology-focused imagery to clearer visualizations of their ‘auxiliary AI’ ideas."

\begin{figure*}[htbp]
    \centering
    \includegraphics[width=0.8\linewidth]{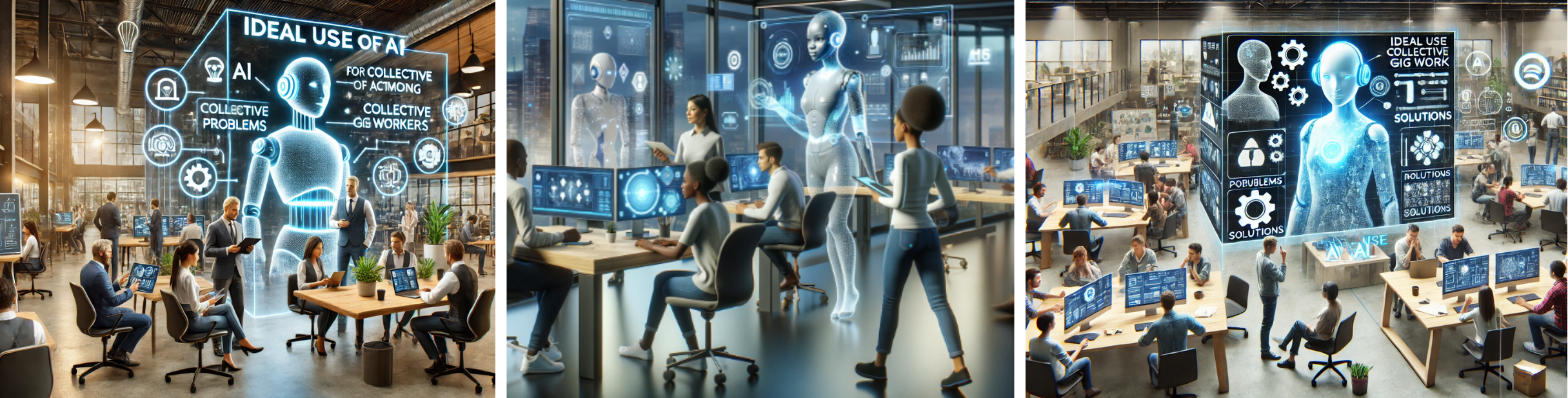}
    \caption{DALL-E generated visions of collaborative AI from three different co-design groups. Left (Group 1): AI as holographic assistant supporting collective problem-solving. Center (Group 2): Transparent AI figures providing background support while humans lead decision-making. Right (Group 6): AI as informational resource on displays while freelancers engage in co-working.}
    \Description{Three DALL-E generated images from different co-design groups showing convergent visions of auxiliary AI. Left panel from Group 1 shows freelancers gathered around a table with a translucent holographic AI figure floating above, providing information while humans discuss. Center panel from Group 2 depicts transparent, ghost-like AI figures standing behind seated human collaborators who lead the discussion. Right panel from Group 6 shows freelancers working together at computers with AI-generated content displayed on wall-mounted screens, keeping AI as an informational resource separate from the human workspace.}
    \label{fig:sharedvisionofAuxAI}
\end{figure*}

\begin{figure*}[htbp]
    \centering
    \includegraphics[width=0.8\linewidth]{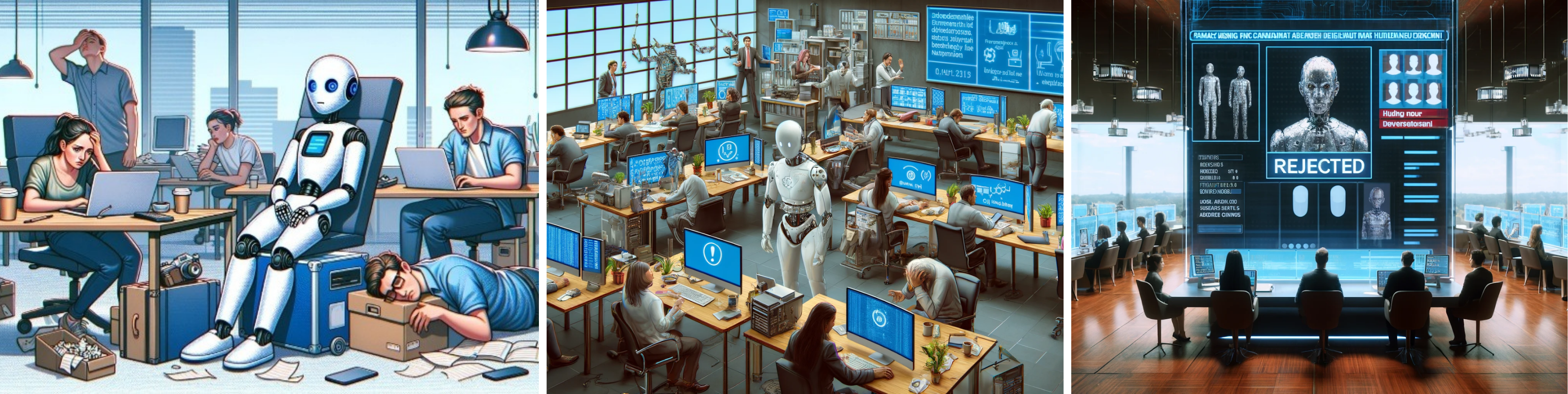}
    \caption{Participant-generated visualizations of undesirable AI futures. Left (Group 3): AI Over-reliance weakening human critical thinking and creativity. Center (Group 4): AI as equal collaborator rather than auxiliary support. Right (Group 7): AI autonomously screening and rejecting potential team members for collaborative projects.}
    \Description{Three DALL-E generated images depicting undesirable AI futures that participants wanted to avoid. Left panel shows a large, prominent robot dominating the workspace while human workers appear disengaged and passive in the background. Center panel depicts a humanoid robot seated at a conference table as an equal team member alongside humans. Right panel shows an AI recruitment system with screens displaying candidate profiles and a prominent red rejected notification, illustrating autonomous AI decision-making without human oversight.}
    \label{fig:UndesriableAI}
\end{figure*}

\subsection{Cross-Group Convergence on Design Principles}
Figure~\ref{fig:sharedvisionofAuxAI} demonstrates that despite working across different co-design sessions, multiple groups (each group had 4-5 freelancers collaboratively refining their design probes) converged on similar visual representations. This convergence suggests that DALL-E served as an effective boundary object, enabling freelancers to articulate shared values about human-led collaboration.

\subsection{Articulating Concerns and Counter-Examples}

Figure~\ref{fig:UndesriableAI} shows how participants used DALL-E to visualize undesirable futures, scenarios they wanted collaborative AI to avoid. This capacity to generate counterexamples proved valuable for articulating design constraints that abstract discussion alone could not achieve.

\end{document}